\documentclass[12pt]{article}

\usepackage[utf8]{inputenc}
\usepackage{geometry}
\usepackage{graphicx}
\usepackage{array}
\usepackage{subfig}
\usepackage{slashed}
\usepackage{amsmath}
\usepackage{amsfonts}
\usepackage{amssymb}
\usepackage[cm]{fullpage}
\usepackage{cite}
\usepackage{epsfig}
\usepackage{comment}
\usepackage{hyperref}
\usepackage{cancel}
\usepackage{mathrsfs}

\usepackage{cleveref}
\crefname{section}{§}{§§}
\Crefname{section}{§}{§§}

\numberwithin{equation}{section}

\def\p{\partial}

\def\0{{(0)}}
\def\1{{(1)}}
\def\2{{(2)}}

\def\<{\langle }
\def\>{\rangle }

\newcommand{\bea}{\begin{eqnarray}}

\newcommand{\eea}{\end{eqnarray}}

\newcommand{\be}{\begin{equation}}
\newcommand{\ee}{\end{equation}}
\newcommand{\ba}{\begin{align}}
\newcommand{\ea}{\end{align}}

   \makeatletter
  \let\over=\@@over \let\overwithdelims=\@@overwithdelims
  \let\atop=\@@atop \let\atopwithdelims=\@@atopwithdelims
  \let\above=\@@above \let\abovewithdelims=\@@abovewithdelims
\renewcommand\section{\@startsection {section}{1}{\z@}%
                                   {-3.5ex \@plus -1ex \@minus -.2ex}%nn
                                   {2.3ex \@plus.2ex}%
                                   {\normalfont\large\bfseries}}

\renewcommand\subsection{\@startsection{subsection}{2}{\z@}%
                                     {-3.25ex\@plus -1ex \@minus -.2ex}%
                                     {1.5ex \@plus .2ex}%
                                     {\normalfont\bfseries}}

\newcommand{\beq}{\begin{equation}}
\newcommand{\eeq}{\end{equation}}
\newcommand{\beqa}{\begin{eqnarray}}
\newcommand{\eeqa}{\end{eqnarray}}
\newcommand{\beqar}{\begin{eqnarray*}}

\newcommand{\mc}[1]{\mathcal{#1}}

\def\[{\[}
\def\]{\]}

\newcommand{\bd}[1]{\begin{fmffile}{#1}\begin{fmfgraph*}}
\newcommand{\ed}{\end{fmfgraph*}\end{fmffile}}

\newcommand{\mN}{\mathcal{N}}
\newcommand{\mb}{\mathbb}

\newcommand{\Nstar}{\mN=2^\star}

\begin{document}

\begin{titlepage}

\begin{flushright}
	CERN-PH-TH/2015-263
\end{flushright}

\unitlength = 1mm~\\
\vskip 1cm
\begin{center}

{\LARGE{\textsc{$\mc{N}=2^\star$ from Topological Amplitudes  \\[0.3cm] in String Theory}}}

\vspace{0.8cm}
Ioannis Florakis\,{}\footnote{\tt ioannis.florakis@cern.ch} and Ahmad Zein Assi\,{}\footnote{\tt azein\_as@ictp.it}

\vspace{1cm}

{\it  ${}^1$ Theory Division - CERN, CH-1211 Geneva 23, Switzerland \\
	${}^2$ High Energy Section - ICTP, Strada Costiera, 11-34014 Trieste, Italy 

}

\vspace{0.8cm}

\begin{abstract}
In this paper, we explicitly construct string theory backgrounds that realise the so-called $\mN=2^\star$ gauge theory. We prove the consistency of our models by calculating their partition function and obtaining the correct gauge theory spectrum. We further provide arguments in favour of the universality of  our construction which covers a wide class of models all of which engineer the same gauge theory. We reproduce the corresponding Nekrasov partition function once the $\Omega$-deformation is included and the appropriate field theory limit taken. This is achieved by calculating the topological amplitudes $F_g$ in the string models. In addition to heterotic and type II constructions, we also realise the mass deformation in type I theory, thus leading to a natural way of uplifting the result to the instanton sector.

\end{abstract}

\vspace{1.0cm}

\end{center}

\end{titlepage}

\pagestyle{empty}
\pagestyle{plain}

\def\vx{{\vec x}}
\def\p{\partial}
\def\po{$\cal P_O$}

\pagenumbering{arabic}

\tableofcontents
\bibliographystyle{utphys}

\section{Introduction}

During the last decades, the interplay between string theory and supersymmetric gauge theories has been the driving force for many discoveries in both fields. One of the most striking examples is the connection between topological string theory \cite{TopoString} and the $\mN=2$ gauge theory with the $\Omega$-deformation \cite{Losev:1997bz,Nekrasov:2002qd}. Indeed, it has been realised that the partition function of the topological string reduces, in the field theory limit, to the free energy of the $\Omega$-deformed $\mN=2$ gauge theory, often referred to as the Nekrasov partition function. This correspondence is valid in the so-called topological limit of the $\Omega$-background, in which one of its parameters is set to zero, while the other is identified with the topological string coupling $g_{\rm s}$. Therefore, what plays the role of a regularisation parameter in  gauge theory acquires physical significance once uplifted to string theory. The extension of this connection to the general setup including two deformation parameters is a programme called \emph{refinement} and has led to many fruitful discoveries \cite{Hollowood,Topvertex,MatrixTheory,Hellerman:2011mv}. In particular, a worldsheet realisation of the $\Omega$-background has been carried out in \cite{AFHNZ,Antoniadis:2013mna}, even though the explicit definition of the twisted, topological theory is still lacking.

From  the point of view of the physical string in this approach, the $\Omega$-deformation boils down to a background of anti-self-dual graviphotons \cite{Bershadsky:1993cx,Antoniadis:1993ze} and self-dual gauge field strenghts \cite{AFHNZ,Antoniadis:2013mna}. The gauge theory partition function descends from a class of BPS amplitudes which has been studied both in the $\mN=2$ \cite{Ooguri:1995cp,Antoniadis:1996qg,Antoniadis:2009nv,Antoniadis:2010iq} as well as in the $\mN=4$ case \cite{Antoniadis:2006mr,Antoniadis:2007cw}. Their BPS nature translates itself into the holomorphic moduli dependence of the corresponding coupling in the string effective action. However, this property is broken at the string level due to boundary effects as expressed, for instance, by the holomorphic anomaly equation \cite{Bershadsky:1993cx}. The generalisation of the latter to the refined case has been analysed in \cite{Huang:2010kf} (see also \cite{Krefl:2010fm}) and more recently in \cite{NewRecRel} from the worldsheet perspective.

In the present work, we study a particular deformation of $\mN=4$ Super Yang-Mills theory, commonly referred to as $\Nstar$. It corresponds to a mass deformation of the former under which the $\mN=2$ adjoint hypermultiplet acquires a mass. Hence, analyticity of the mass parameter renders this theory an interpolation between the pure $\mN=4$ and $\mN=2$ theories, the latter being recovered as particular limits of zero and infinite mass, respectively. In addition, the $\Nstar$ theory is a flagship example in the context of the AGT conjecture \cite{Alday:2009aq} which relates it to the two-dimensional Liouville theory on a torus with one puncture playing the role of the massive hypermultiplet. In this correspondence, the Nekrasov partition function is mapped to the Liouville theory conformal block. More general connections can be established by obtaining four-dimensional gauge theories from the $(2,0)$ theory compactified on a genus $g$ Riemann surface with $n$ punctures \cite{Gaiotto:2009we}, and considering Liouville theory on the Riemann surface. For instance, a torus with four punctures leads to the $\mN=2$, $\rm SU(2)$ gauge theory with four flavours.

Since a worldsheet description of $\Nstar$ in string theory  is lacking, our goal is to fill this gap by studying string theory with spontaneous breaking of supersymmetry from $\mN=4$ to $\mN=2$, in such a way that the adjoint hypermultiplet acquires a moduli dependent mass. This is achieved by considering a freely-acting orbifold of the $T^6$ torus, implementing the uplift of the Scherk-Schwarz mechanism to string theory \cite{Ferrara:1987qp,Kounnas:1988ye,Ferrara:1988jx,Kiritsis:1997ca,Antoniadis:1998ep,Condeescu:2012sp}. In this construction, certain  states acquire a mass that is inversely proportional to the volume of some cycle of the internal space and the unbroken supersymmetric model is recovered in the limit of zero mass. We first present a heterotic asymmetric orbifold construction which turns out to be the most natural way of realising $\mN=2^\star$ in perturbative closed string theory. We show that it indeed  provides mass to the adjoint hypermultiplet and correctly reproduces the $\Nstar$ spectrum without modifying the gauge group. More generally, we present a correspondence between a wide class of freely-acting orbifolds of $\mN=4$ compactifications and the $\Nstar$ gauge theory. We elaborate on the universality of this construction and provide evidence supporting the correspondence. In particular, we compute topological amplitudes for these $\Nstar$ theories, for symmetric and asymmetric freely acting orbifolds, and confirm that they correctly reduce, in the field theory limit, to the $\Omega$-deformed partition function \cite{Nekrasov:2003rj,Billo:2013fi,Pestun:2007rz} of the $\Nstar$ gauge theory. This class of BPS amplitudes, which has proven useful in the study of string dualities, is also analysed in type I theory. The latter turns out to be a natural framework for incorporating gauge theory instanton corrections.

The paper is structured as follows. In Section \ref{AsymOrb}, we begin by explicitly constructing a worldsheet realisation of $\Nstar$ as an asymmetric freely-acting orbifold in heterotic string theory, analyse its string partition function and show the accordance with the spectrum of the $\Nstar$ gauge theory. We further confirm our result by calculating the topological couplings $F_g$ at the one-loop level in string perturbation theory and show that they correctly reproduce the perturbative part of the mass-deformed Nekrasov partition function. In Section \ref {StringModel}, we formulate a general correspondence between certain classes of freely-acting orbifolds in string theory and the $\Nstar$ gauge theory, and illustrate our construction in terms of an explicit symmetric orbifold model with a type IIA dual. Moreover, in Section \ref{TypeOne}, we realise the mass deformation in type I orientifolds and derive the Nekrasov partition function in a specific model.  Finally, Section \ref{Conclusion} summarises our conclusions. 

%%%

%%%%%%%%%%%%%%%%%%%%%%%%%%%%%%%%

\section{\texorpdfstring{$\Nstar$ realisation in string theory}{N=2 star realisation in string theory}}\label{AsymOrb}

\subsection{An asymmetric orbifold model}

In order to realise the $\mN=2^\star$ theory in the heterotic string, we begin with the ${\rm E}_8\times {\rm E}_8$ heterotic model\footnote{A similar asymmetric construction can also be performed for the ${\rm SO}(32)$ heterotic string.} on $T^6$ and identify a suitable orbifold action that gives mass to part of the $\mN=4$ vector multiplet and to two gravitini. As shown below, the resulting string theory has eight unbroken supercharges and a massive adjoint hypermultiplet. In order for the orbifold not to project out non-invariant states but rather give them non-trivial masses, the orbifold in question must act freely on the $T^6$ coordinates, {\it i.e.} without fixed points. The simplest possibility is to consider a $\mathbb Z_2$ orbifold which couples rotations on two of the complexified coordinates parametrising the $T^6$ with a translation along the third complexified coordinate. Hence, we obtain a freely-acting variation of the standard orbifold realisation of the K3 surface as $T^4/\mathbb Z_2$. 

To identify the orbifold action, we start by constructing the vertex operators of the fields in the $\mN=4$ vector multiplet. For simplicity, we work in the light-cone gauge. Let us denote by $Z^1$, $Z^2$, $Z^3$ the complexified coordinates on $T^6$ and by $\Psi^1$, $\Psi^2$, $\Psi^3$ their fermionic worldsheet superpartners. The latter can be bosonised in terms of three chiral bosons $\Phi^j(z)$ as
\begin{equation}
	\Psi^j(z) = e^{i\sqrt{2}\Phi^j} \ ,\qquad \bar\Psi^j(z) = e^{-i\sqrt{2}\Phi^j} \ ,\qquad j=1,2,3 \ ,
\end{equation}
where we have conventionally set  $\alpha'=1$. In order for the resulting theory to enjoy an unbroken $\mN=2$ supersymmetry, the $\mathbb Z_2$ orbifold must rotate the fermionic coordinates in the planes $j=2,3$ with opposite angles $\pi$, $-\pi$
\begin{equation}
	\begin{split}
	&\Psi^2 \to e^{i\pi} \Psi^2  \ ,\qquad ~\,  \Phi^2 \to \Phi^2 +\tfrac{\pi}{\sqrt{2}} \,, \\
	&\Psi^3 \to e^{-i\pi} \Psi^3 \ , \qquad  \Phi^3 \to \Phi^3 - \tfrac{\pi}{\sqrt{2}} \,.
	\end{split}
\label{orbFerm}
\end{equation}
The presence of the $\sqrt{2}$ factors in the above expressions is consistent with the fact that bosonisation of the worldsheet fermions occurs at the fermionic radius $r=1/\sqrt{2}$. Note also that the above action on the fermions induces the breaking of the ${\rm SO}(6)$ R-symmetry group of $\mN=4$ down to ${\rm SO}(2)\times {\rm SO}(4)$.

The worldsheet fermions $\psi^\mu$ of the ten-dimensional theory transform under the ${\rm SO}(8)$ little group. Upon toroidally compactifying down to four dimensions, one may decompose ${\rm SO}(8)\to {\rm SO}(4)\times {\rm SO}(4)$, where the first ${\rm SO}(4)$ factor corresponds to the transverse spacetime fermions $\psi^\mu$ together with the complex fermion $\Psi^1$ associated to the internal $T^2$ that is left unrotated by the $\mathbb Z_2$ orbifold. The second ${\rm SO}(4)$ factor then corresponds to the complex fermions $\Psi^2$, $\Psi^3$ in the $T^4$ directions that are rotated by the orbifold action.
Similarly, we define spin fields for the fermions associated to the two ${\rm SO}(4)$ factors via
\begin{equation}
	\begin{split}
	S_{\alpha} &= e^{\pm\frac{i}{\sqrt{2}}( \Phi^0+\Phi^1)}\ ,  \,\qquad S_{\dot\alpha} = e^{\pm\frac{i}{\sqrt{2}}( \Phi^0-\Phi^1)} \,,\\
	\Sigma_{A} &= e^{\pm\frac{i}{\sqrt{2}}( \Phi^2+\Phi^3)}\ ,\qquad \Sigma_{\dot A} = e^{\pm\frac{i}{\sqrt{2}}( \Phi^2-\Phi^3)} \,.
	\end{split}
\end{equation}
The indices $\alpha, \dot\alpha$ correspond to the (Weyl) spinor and conjugate spinor representations of the first ${\rm SO}(4)$ factor and, similarly,  $A,\dot A$ label the two Weyl spinor representations associated to the second ${\rm SO}(4)$ factor.

In the  above notation, the $\mN=4$ vector multiplet vertex operators may be written explicitly as
\begin{equation}
	\begin{split}
		{\rm a vector\,boson} \quad  &: \quad \psi^\mu(z) \, \bar J(\bar z)  \,, \\
		{\rm gaugini} \quad  &: \quad  S_{\dot\alpha}\,\Sigma_A(z)\, \bar J(\bar z) \ ,\quad S_\alpha \, \Sigma_{\dot A}(z) \, \bar J(\bar z) \, ,\\
		{\rm scalars} \quad  &: \quad e^{\pm i\sqrt{2}\Phi^j(z)} \, \bar J(\bar z)\,,  \qquad\ j=1,2,3 \,,
	\end{split}
\end{equation}
where $\bar J(\bar z)$ is a right-moving current in the Kac Moody algebra of the gauge group. It is now straightforward to see that under the $\mathbb Z_2$ action  \eqref{orbFerm} on the fermions, the left-moving operators 
\begin{equation}
	S_\alpha\,\Sigma_{\dot A}(z) \ ,\qquad e^{\pm i\sqrt{2}\Phi^2} \ ,\qquad e^{\pm i\sqrt{2} \Phi^3} \,,
\end{equation}
transform with a minus sign, which implies the decomposition of the $\mN=4$ vector multiplet into an $\mN=2$ vector multiplet:
\begin{equation}
	\begin{split}
		{\rm vector\,boson} \quad  &: \quad \psi^\mu(z) \, \bar J(\bar z)  \,, \\
		{\rm gaugini} \quad  &: \quad  S_{\dot\alpha}\,\Sigma_A(z)\, \bar J(\bar z) \, ,\\
		{\rm scalars} \quad  &: \quad e^{\pm i\sqrt{2} \Phi^1(z)} \, \bar J(\bar z)  \,,
	\end{split}
\end{equation}
and an $\mN=2$ adjoint hypermultiplet 
\begin{equation}
	\begin{split}
		{\rm fermion} \quad  &:  \quad S_\alpha \, \Sigma_{\dot A}(z) \, \bar J(\bar z) \, ,\\
		{\rm scalars} \quad  &: \quad e^{\pm i\sqrt{2}\Phi^j(z)} \, \bar J(\bar z) \ , \qquad  j=2,3 \,.
	\end{split}
\end{equation}
In order to preserve an unbroken local $\mN=(1,0)$ worldsheet supersymmetry, the action \eqref{orbFerm} of the $\mathbb Z_2$ orbifold on the fermions needs to be supplemented with the corresponding $\mathbb Z_2$ action on the left-moving worldsheet bosons
\begin{equation}
	Z_L^2 \to e^{i\pi}\, Z_L^2 \ ,\qquad Z_L^3\to e^{-i\pi}\,Z_L^3 \,.
\label{orbBos}
\end{equation}
If the $\mathbb Z_2$ rotation is consistently coupled to a translation along some direction in the first plane $Z^1$, states that are charged under the action of the rotation are no longer projected out in the orbifolded theory, but rather acquire a Scherk-Schwarz mass proportional to their charge.

 It is then straightforward to see that in order to render the $\mN=2$ hypermultiplet massive via the Scherk-Schwarz mechanism without affecting the gauge group, the orbifold action on the right moving currents $\bar J$ must be trivial. In particular, this requires that  the right moving $T^6$ coordinates $Z^i_R(\bar z)$ remain unrotated and the $\mathbb Z_2$ orbifold is identified as an asymmetric, freely acting orbifold generated by
 \begin{equation}
		g = e^{i\pi Q_L}\,\delta \,.
 \end{equation}
Here, $\exp(i\pi Q_L)$ is the operator rotating the worldsheet super-coordinates in the $j=2,3$ complex planes by opposite angles $\pm\pi$ as in eqs. \eqref{orbFerm},\eqref{orbBos}, and $\delta$ is an order two (momentum) shift along the first cycle of the $j=1$ complex plane
\begin{equation}
	\delta\ : \ Z^1\to Z^1+i\pi \sqrt{\tfrac{T_2}{U_2}} \,,
\label{shiftdelta}
\end{equation}
where $T,U$ are the K\"ahler and complex structure moduli of the $T^2$ parametrised by $Z^1=X^2+iX^1$. In the special case of a rectangular 2-torus without $B$ field, corresponding to the factorisation limit $T^2=S^1\times S^1$, the above shift is nothing but a translation by half the circumference of the first circle $X^1$. 

\subsubsection{Spectrum of the theory}

The one loop partition function of the theory reads
\begin{equation}
	Z=\frac{1}{2^2\,\eta^{12}\,\bar\eta^{24}}\sum_{\genfrac{}{}{0pt}{}{a,b=0,1}{H,G=0,1}}(-1)^{a+b+ab+HG}\,\vartheta\bigr[^{a}_{b}\bigr]^2 \vartheta\bigr[^{a+H}_{b+G}\bigr]\vartheta\bigr[^{a-H}_{b-G}\bigr]\,\Gamma_{4,4}\bigr[^H_G\bigr]\,\Gamma_{2,2}\bigr[^H_G\bigr]\,\Gamma_{{\rm E}_8}^2(\bar\tau) \,,
\label{partFunc}
\end{equation}
where $a,b$ are summed over the spin structures on the worldsheet torus,  $H$ labels the orbifold sectors and the sum over $G$ imposes the $\mathbb Z_2$ invariance projection. $\Gamma_{4,4}\bigr[^H_G\bigr]$ is the asymmetrically twisted lattice partition function of the $j=2,3$ planes, $\Gamma_{2,2}\bigr[^H_G\bigr](T,U)$ is the shifted Narain lattice  associated to the $j=1$ plane and $\Gamma_{\rm E_8}(\bar\tau)$ is the anti-chiral $\rm E_8$ lattice partition function.  $\vartheta\bigr[^a_b\bigr](\tau)$ stands for the one-loop Jacobi theta constant with characteristics and $\eta(\tau)$ is the Dedekind eta function.

Due to the asymmetric nature of the orbifold rotation, the K3 lattice is taken to lie at the fermionic factorised point, where the $T^4$ parametrised by $Z^2$, $Z^3$ can be viewed as the product of four circles with all radii equal to $r=1/\sqrt{2}$ and its moduli are stabilised at the minimum of the tree level scalar potential generated by the Scherk-Schwarz gauging of $\mN=4$ supergravity:
\begin{equation}
	V(\chi_\alpha)\sim \frac{1}{S_2 T_2 U_2}\sum_{\alpha} \frac{|1+\chi_{\alpha}^2|^2}{ {\rm Im}(\chi_{\alpha})} \,,
\end{equation}
where $S_2$ is the imaginary part of the heterotic dilaton and $\chi_\alpha$ are the moduli of $T^4$.

 As a result, its partition function can be entirely expressed in terms of one-loop Jacobi theta constants
\begin{equation}
	\Gamma_{4,4}\bigr[^H_G\bigr] = \frac{1}{2}\sum_{\gamma,\delta=0,1}\vartheta\bigr[^\gamma_\delta\bigr]^2\,\vartheta\bigr[^{\gamma+H}_{\delta+G}\bigr]\,\vartheta\bigr[^{\gamma-H}_{\delta-G}\bigr]\,\bar\vartheta\bigr[^\gamma_\delta\bigr]^4 \,.
\end{equation}
On the other hand, the $T,U$ moduli of the $T^2$ parametrised by $Z^1$ remain massless and correspond to the no-scale moduli of the partial supersymmetry breaking. The contribution of the shifted Narain lattice reads
\begin{equation}
	\Gamma_{2,2}\bigr[^H_G\bigr](T,U) = \tau_2 \sum_{m,n} (-1)^{m_1 G}\, q^{\frac{1}{4}|P_L|^2}\,\bar q^{\frac{1}{4}|P_R|^2} \,,
\end{equation}
with $P_L$, $P_R$ being the complexified lattice momenta
\begin{equation}
	P_L = \frac{m_2-\bar U m_1 + \bar T (n^1+\frac{H}{2}+\bar U n^2)}{\sqrt{T_2 U_2}} \ ,\qquad P_R = \frac{m_2-\bar U m_1 +  T (n^1+\frac{H}{2}+\bar U n^2)}{\sqrt{T_2 U_2}} \,.
\end{equation}
In particular, $P_L$ is identified as the central charge of the $\mN=2$ superalgebra.

The one-loop partition function of the theory is consistent with unitarity and modular invariance at all genera, as guaranteed by the theorems of fermionic \cite{Antoniadis:1986rn} and asymmetric orbifold \cite{Narain:1986qm,Narain:1990mw} constructions. It is convenient to rewrite \eqref{partFunc} in terms of ${\rm SO}(2n)$ characters
\begin{equation}
	\begin{split}
	& O_{2n} = \frac{1}{2\eta^n}(\vartheta_3^n+\vartheta_4^n) \ , \qquad \quad  \  V_{2n} = \frac{1}{2\eta^n}(\vartheta_3^n-\vartheta_4^n) \,, \\
	& S_{2n} = \frac{1}{2\eta^n}(\vartheta_2^n+i^{-n}\vartheta_1^n) \ , \qquad C_{2n} = \frac{1}{2\eta^n}(\vartheta_2^n-i^{-n}\vartheta_1^n) \,,
	\end{split}
\end{equation}
from which the spectrum may be easily read. The expansion explicitly reads
\begin{equation}
	\begin{split}
	\eta^2\bar\eta^2\,Z =&(V_4 O_4-S_4 S_4)\bigr(\Gamma_{4,4}\bigr[^{\,0}_+\bigr]\Gamma_{2,2}\bigr[^{\,0}_+\bigr]+\Gamma_{4,4}\bigr[^{\,0}_-\bigr]\Gamma_{2,2}\bigr[^{\,0}_-\bigr]\bigr)\,(\bar O_{16}+\bar C_{16})^2  \\
	+&(O_4 V_4-C_4 C_4)\bigr(\Gamma_{4,4}\bigr[^{\,0}_-\bigr]\Gamma_{2,2}\bigr[^{\,0}_+\bigr]+\Gamma_{4,4}\bigr[^{\,0}_+\bigr]\Gamma_{2,2}\bigr[^{\,0}_-\bigr]\bigr)\,(\bar O_{16}+\bar C_{16})^2  \\
	+&(V_4 C_4-S_4 V_4)\bigr(\Gamma_{4,4}\bigr[^{\,1}_+\bigr]\Gamma_{2,2}\bigr[^{\,1}_+\bigr]+\Gamma_{4,4}\bigr[^{\,1}_-\bigr]\Gamma_{2,2}\bigr[^{\,1}_-\bigr]\bigr)\,(\bar O_{16}+\bar C_{16})^2  \\
	+&(O_4 S_4-C_4 O_4)\bigr(\Gamma_{4,4}\bigr[^{\,1}_+\bigr]\Gamma_{2,2}\bigr[^{\,1}_-\bigr]+\Gamma_{4,4}\bigr[^{\,1}_-\bigr]\Gamma_{2,2}\bigr[^{\,1}_+\bigr]\bigr)\,(\bar O_{16}+\bar C_{16})^2  \,,
	\end{split}
\label{partFuncChar}
\end{equation}
where we decomposed the lattices into irreducible representations
\begin{equation}
	\Gamma_{d,d}\bigr[^H_{\pm}\bigr]=\frac{1}{2\eta^d\bar\eta^d}\,\bigr( \Gamma_{d,d}\bigr[^H_{\,0}\bigr]\pm \Gamma_{d,d}\bigr[^H_{\,1}\bigr] \bigr) \,,
\end{equation}
so that the $\pm$ sign indicates the eigenvalue of the object with respect to the $\mathbb Z_2$ action, and we included also the associated oscillator contributions encoded in the Dedekind functions. 

It is clear that the shifted lattice with $H=1$ carries non-trivial winding charge and the corresponding states decouple in the low energy limit. We hence focus our attention on the untwisted sector $H=0$. In this sector, the $\mathbb Z_2$ invariant shifted lattice $\Gamma_{2,2}\bigr[^{\,0}_{+}\bigr]$  projects to the $m_1=0\,{\rm mod}\, 2$ subsector and gives rise to massless states with vanishing momentum and winding numbers. On the other hand, the $\mathbb Z_2$ charged lattice $\Gamma_{2,2}\bigr[^{\,0}_{-}\bigr]$ instead projects to the $m_1=1\,{\rm mod}\,2$ subsector and is responsible for attributing  Scherk-Schwarz masses to charged states.

The twisted lattice contribution to the conformal weights can be read off from its explicit expansion around the cusp $\tau=\infty$:
\begin{equation}
	\begin{split}
	&\Gamma_{4,4}\bigr[^{\,0}_{+}\bigr] = (q\bar q)^{1/6}\left(1 +12\,q+28\, \bar q+96\, q^{1/2}\,\bar q^{1/2}+\ldots\right) \,, \\
	&\Gamma_{4,4}\bigr[^{\,0}_{-}\bigr] = (q\bar q)^{1/6}\left(96\,q^{1/2}\,\bar q^{1/2}+16\,q+\ldots\right) \,.
	\end{split}
\end{equation}
Notice, in particular, that the non-invariant combination $\Gamma_{4,4}\bigr[^{\,0}_{-}\bigr]$ always contributes at least $1/2$ unit of left-moving conformal weight, hence giving rise to states with string scale masses, which can be similarly ignored in a low energy analysis of the spectrum.

The $\mN=2$ vector multiplets arise from the $V_4 O_4-S_4 S_4$ part in eq. \eqref{partFuncChar} and the matter gauge group of the theory at a generic point in the $T,U$ moduli space is identified as ${\rm G}={\rm E}_8\times {\rm E}_8\times {\rm SO}(8)\times {\rm U}(1)^2$. The $\rm U(1)$ factors arise from the dimensional reduction on the unrotated $T^2$ parametrised by $Z^1$ and correspond to the currents $\bar J(\bar z) = i\bar\partial X^1$ or $i\bar\partial X^2$. On the other hand, the ${\rm SO}(8)$ factor is a consequence of the enhancement ${\rm U}(1)^4\to {\rm SO}(8)$, due to the compactification of the $T^4$ directions $Z^2=X^4+iX^3$, $Z^3=X^6+iX^5$  at the fermionic radii. The ${\rm SO}(8)$ Kac Moody algebra is then realised by the currents
\begin{equation}
	\bar J(\bar z) =  \{ \ i\bar\partial X^a \ , \quad e^{\pm i\sqrt{2}X_R^b(\bar z) \pm i\sqrt{2} X_R^c(\bar z)} \ \} \ ,\quad  b\neq c \,,
\end{equation}
with $a,b,c=1,\ldots, 4$.

The hypermultiplets  arise from the $O_4 V_4-C_4 C_4$ part of eq. \eqref{partFuncChar} and are massive, by construction. This can be seen by noticing that these states carry odd momentum charge along the direction of the shift.

The low lying states of interest, which become massless in the limit where $\mN=4$ supersymmetry is recovered, are given by $m_1=\pm 1$ and $n^1=0$. The first possibility is to take $m_2=n^2=0$, in which case the corresponding  vertex operators are
\begin{equation}
	\left.\begin{array}{l}
	S_\alpha \,\Sigma_{\dot A} (z) \,\bar J(\bar z)\\ 
	e^{\pm i\sqrt{2}\Phi^j(z)}\,\bar J(\bar z)\\
	\end{array}\right\} \times \exp\left[\pm i\,\frac{UZ^1(z,\bar z)+\bar U \bar Z^1(z,\bar z)}{2\sqrt{T_2 U_2}}\right] \,.
\label{HyperE8}
\end{equation}
Note that they are invariant under the orbifold action, since the shift in $Z^1$ eq. \eqref{shiftdelta} precisely cancels the the minus sign from the left-movers.
Their physical mass is given by
\begin{equation}
	m_{\rm hyp,1}^2 = \frac{|U|^2}{T_2 U_2} \,,
\end{equation}
and is further identified as the scale $m_{3/2}^2$ of the  $\mN=4\to \mN=2$ supersymmetry breaking. The presence of $\bar J$ in the vertex operators ensures that the hypers  indeed transform in the adjoint of the gauge group ${\rm G}$, as expected. 

The second possibility is to consider $m_2=n^2=\pm 1$, in which case the corresponding vertex operators are
\begin{equation}
	\left.\begin{array}{l}
	S_\alpha \,\Sigma_{\dot A} (z) \\ 
	e^{\pm i\sqrt{2}\Phi^j(z)}\\
	\end{array}\right\} \times \exp\left[ \frac{i}{2}\bigr( P_L \bar Z_L^1+\bar P_L Z_L^1+P_R \bar Z_R^1+\bar P_R Z_R^1\bigr)\right] \,,
	\label{Hyper2}
\end{equation}
with the lattice momenta given by
\begin{equation}
	P_L=\frac{\pm(1+\bar T\bar U)\pm \bar U}{\sqrt{T_2 U_2}}\ , \quad P_R = \frac{\pm(1+T\bar U)\pm\bar U}{\sqrt{T_2 U_2}} \,.
\label{HyperSU2}
\end{equation}
These states are again massive, with their physical mass being given by
\begin{equation}
	m^2_{\rm hyp,2} = \frac{|1+TU\pm U|^2}{T_2 U_2} \,.
\end{equation}
They correspond to states charged under the adjoint of an $\rm SU(2)$ gauge group that arises as a result of enhancement, as we discuss below.

\subsubsection{Enhanced gauge symmetry}

We next investigate the possibility of enhanced gauge symmetry at special points in the moduli space ${\rm SO}(2,2)/{\rm SO}(2)\times{\rm SO}(2)$ parametrised by the $T,U$ moduli of the unrotated $T^2$ torus. Clearly, this can only arise from the sector
\begin{equation}
	(V_4 O_4-S_4 S_4)\,\Gamma_{2,2}\bigr[^{\,0}_{+}\bigr] \,.
\end{equation}
In order to create a level matched state, we demand that the lattice momenta satisfy $|P_L|^2-|P_R|^2=-4(m_1 n^1+m_2 n^2)=-4$. Since in this sector $m_1=0\,{\rm mod}\,2$, it is sufficient to choose the state $m_1=n^1=0$ and $m_2=n^2=\pm 1$, in which case the vertex operators of the extra charged states read
\begin{equation}
	\left.\begin{array}{l}
	\psi^\mu(z) \\
	S_{\dot\alpha} \,\Sigma_{ A} (z)\\ 
	e^{\pm i\sqrt{2}\Phi^1(z)}\\
	\end{array}\right\} \times \exp\left[ \frac{i}{2}\bigr( P_L \bar Z_L^1+\bar P_L Z_L^1+P_R \bar Z_R^1+\bar P_R Z_R^1\bigr)\right] \,,
\end{equation}
where
\begin{equation}
	P_L = \pm \frac{1+\bar T\bar U}{\sqrt{T_2 U_2}}\ , \qquad P_R = \pm \frac{1+T\bar U}{\sqrt{T_2 U_2}} \,.
\end{equation}
and their physical mass is
\begin{equation}
	m^2_{\rm vec} = \frac{|1+TU|^2}{T_2 U_2} \,.
\end{equation}
In particular, they become massless at the point $T=-1/U$. Note that, contrary to the $\mN=2$ string theory obtained by compactifying the heterotic string on ${\rm K}3\times T^2$, in the present $\mN=2^\star$ case, the T-duality group is no longer the full ${\rm SL}(2;\mathbb Z)_T \times {\rm SL}(2;\mathbb Z)_U \ltimes \mathbb Z_2$ but is actually broken to its subgroup $\Gamma^0(2)_T\times \Gamma_0(2)_U\ltimes\mathbb Z_2$. As a result, this point of gauge symmetry enhancement is not equivalent to the $T=U$ one, since $\binom{0~-1}{1~~0}\notin\Gamma_0(2)$. This is an effect originating from the freely acting nature of the $\mathbb Z_2$ orbifold.

At the enhancement point $T=-1/U$, the ${\rm U}(1)\times {\rm U}(1)$ gauge symmetry becomes enhanced to ${\rm SU}(2)\times {\rm U}(1)$. Explicitly, the vertex operators of the massless vector multiplets transforming under the ${\rm SU}(2)\times {\rm U}(1)$ become
\begin{equation}
	\left.\begin{array}{l}
	\psi^\mu(z) \\
	S_{\dot\alpha} \,\Sigma_{ A} (z)\\ 
	e^{\pm i\sqrt{2}\Phi^1(z)}\\
	\end{array}\right\} \times \{ \ \bar J_0(\bar z)\ , \  \bar J_{\pm}(\bar z)\ , \  \hat{\bar J}_0(\bar z) \ \} \,,
\end{equation}
where
\begin{equation}
	\bar J_0(\bar z) = i\, \frac{U_1\,\bar\partial X^1+U_2\,\bar\partial X^2}{|U|} \ ,\qquad \bar J_\pm(\bar z) = \exp\left[\pm 2i\,\frac{U_1 X_R^1(\bar z)+U_2 X_R^2(\bar z)}{|U|}\right] \,,
\end{equation}
realise the ${\rm SU}(2)$ Kac Moody algebra at level $k=1$, whereas the remaining ${\rm U}(1)$ is generated by the orthogonal combination
\begin{equation}
	\hat{\bar J}_0(\bar z) = i\, \frac{U_2\,\bar\partial X^1-U_1\,\bar\partial X^2}{|U|} \,.
\end{equation}
This U(1) is the one used in the Scherk-Schwarz mechanism in order to give masses to all  hypermultiplets of the theory. 

At the $T=-1/U$ enhancement point, the hypermultiplets transforming in the adjoint of ${\rm E}_8\times{\rm E}_8\times {\rm SO}(8)$ and those transforming in the adjoint of ${\rm SU}(2)$ given in eqs.\eqref{HyperE8} and \eqref{HyperSU2}, respectively, become degenerate in mass, $m^2_{\rm hyp,1}=m^2_{\rm hyp,2}=|U|^4/U_2^2$. This degeneracy can be lifted by turning on non-trivial Wilson lines along the unrotated $T^2$ for the ${\rm E}_8\times{\rm E_8}\times{\rm SO}(8)$ gauge group factors, breaking them to ${\rm U}(1)^{24}$. The latter can be accomplished by adding a marginal deformation of the form $A_i^a\, \partial X^i\bar J^a(\bar z)$ to the worldsheet Lagrangian. The effect of the Wilson line is to modify the masses of the states so that the above enhancement point is now modified to $TU-W^2/2=-1$. The hypermultiplet masses are also modified:
\begin{equation}
	m^2_{\rm hyp,1}= \frac{|Q\cdot W+U|^2}{T_2 U_2-\frac{1}{2}W_2^2}\ ,\quad m^2_{\rm hyp,2}=\frac{|1+TU-\frac{1}{2}W^2\pm U|^2}{T_2 U_2-\frac{1}{2}W_2^2} \,,
\end{equation}
with $Q^a$ being the associated Cartan charges associated to the roots of ${\rm E}_8\times{\rm E}_8\times{\rm SO}(8)$, normalised to length $Q^2=2$, and $W^a=Y^a_2-U Y^a_1$ being the complexified Wilson line. Notice that, indeed, at the $TU-W^2/2=-1$ enhancement point, the hypermultiplet masses $m^2_{\rm hyp,1}$ with $Q\neq 0$ and $m^2_{\rm hyp,2}$ are no longer degenerate. This hierarchy of masses permits one to focus on the field theory limit of various gauge group factors in ${\rm G}$ by consistently decoupling the degrees of freedom of the others. For simplicity, we henceforth focus on the ${\rm SU}(2)$ gauge factor. The case $Q=0$, on the other hand, corresponds to the case where $\bar J$ in \eqref{HyperE8} is associated with the Cartan current $\bar J_0$ of SU(2). At the enhancement point, this state becomes degenerate with the hypermultiplets \eqref{Hyper2} of mass $m_{\rm hyp,2}^2$ and completes the adjoint representation of SU(2).

%%%%%

\subsection{\texorpdfstring{Topological amplitudes in $\mN=2^\star$}{Topological amplitude in N=2*}}\label{TopoAmp}

A well-defined check of our proposal for the realisation of $\Nstar$ in terms of the asymmetric orbifold model discussed in the previous section is the computation of the partition function of the mass deformed gauge theory on $\mathbb R^4\times S^1$ twisted by equivariant rotation parameters $\epsilon_{1,2}$. More precisely, exploit the relation of the Nekrasov partition function of gauge theory to the free energy of topological string theory or, equivalently, to physical string amplitudes involving insertions of anti-self-dual graviphoton field strengths. Indeed,  the field theory limit of the genus $g$  partition function $F_g$  of topological string theory on a Calabi-Yau manifold, is related to the Nekrasov partition function $Z_{\rm Nek}$ via
\begin{align}
	\sum_{g\geq 0}\epsilon^{2g-2}\,F_g \Bigr|_{\rm F.T.}= \log\, Z_{\rm Nek}(\epsilon_1=-\epsilon_2\equiv \epsilon)\,.
\end{align}
On the other hand, $F_g$ computes physical, genus $g$ string amplitudes of the (untwisted) type II theory, involving two insertions of anti-self-dual Riemann tensors and $2g-2$ insertions of anti-self-dual graviphoton field strengths. In the dual heterotic theory, these couplings start receiving corrections at genus one while being protected against higher perturbative corrections. As a result, in order to extract the perturbative part of the gauge theory partition function corresponding to our asymmetric orbifold model, it is sufficient to compute heterotic topological amplitudes at genus one and expand their field theory limit around a suitable point of enhanced gauge symmetry. Specifically, we focus on the SU(2) enhancement arising at the $TU-W^2/2=-1$ point.

\subsubsection{Setup and derivation of the amplitude}

The topological amplitudes of interest calculate higher derivative terms in the string effective action of the form
\begin{equation}
	\int d^4\theta\, F_g(X)\,W^{2g} = F_g(\phi)\,R^2_{-}\,F^{2g-2}_{-}+\ldots\,.
	\label{ActionTopAmpl}
\end{equation}
Here, $R_{-}$, $F_{-}$ are the anti self-dual Riemann tensor and graviphoton field strength, respectively, which arise as the bosonic components of the Weyl superfield $W$. Furthermore, $X$ collectively denotes the chiral superfields of the theory with scalar component $\phi$.

In heterotic string perturbation theory, these amplitudes receive contributions at one loop in $g_{\rm s}$ and are otherwise protected against higher perturbative corrections. Denoting by $Z^4, Z^5$ the complexified spacetime coordinates and by $\Psi^4,\Psi^5$ their fermionic superpartners, one may write the relevant vertex operators in a convenient kinematic configuration as
\begin{equation}
	\begin{split}
	V_{ R_-}(p_4) = (\partial Z^5-ip_4\Psi^4\Psi^5)\,\bar\partial Z^5\,e^{ip_4 \,Z^4}\,,\\
	V_{ R_-}(\bar p_4) = (\partial \bar Z^5-i\bar p_4\bar\Psi^4\bar\Psi^5)\,\bar\partial \bar Z^5\,e^{i\bar p_4 \,\bar Z^4}\,,\\
	V_{ F_-}(p_4) = (\partial Z^1-ip_4\Psi^4\Psi^1)\,\bar\partial Z^5\,e^{ip_4 \,Z^4}\,,\\
	V_{ F_-}(\bar p_4) = (\partial Z^1-i\bar p_4\bar\Psi^4\Psi^1)\,\bar\partial \bar Z^5\,e^{i\bar p_4 \,\bar Z^4}\,.
	\end{split}
\end{equation}
In calculating the correlator $\langle R^2_{-}F_{-}^{2g-2}\rangle$, one notices that the effect of the two Riemann tensor insertions is simply to soak up the $\Psi^4,\Psi^5$ zero modes, as reflected in \eqref{ActionTopAmpl}. In addition, the worldsheet fermions $\Psi^1$ in the graviphoton vertex operators do not contract and their one loop contribution cancels against that of the superghost. The correlator reads
\begin{equation}
	(p_4\bar p_4)^{g-1} \bigr\langle \prod_i^{g-1}\int dx_i \,\partial Z^1 (Z^4\bar\partial Z^5)(x_i)\,\prod_j^{g-1}\int dy_j\,\partial Z^1 (\bar Z^4\bar\partial\bar Z^5)(y_j)\bigr\rangle\,.
\end{equation}
This can be straightforwardly calculated by taking suitable derivatives of the generating function
\begin{equation}
	\mathcal F(\epsilon)=\sum_{g}\frac{\epsilon^{2g}}{(g!)^2}\,F_g 
					= \biggr\langle \exp\Bigr[-\epsilon\int d^2 z\,\partial Z^1(Z^4\bar\partial Z^5+\bar Z^5\bar\partial \bar Z^4) \Bigr] \biggr\rangle \,,
\end{equation}
where $\epsilon$ can be physically viewed as a vacuum expectation value for the graviphoton field strength.
Since $Z^1$ contributes only through its zero mode, the path integral over $Z^4,Z^5$ becomes effectively gaussian and is explicitly evaluated to be
\begin{equation}
	G_{\rm bos}(\epsilon)= \frac{(2\pi i\epsilon)^2 \bar\eta^6}{\bar\vartheta_1(\tilde\epsilon)^2}\,e^{-\frac{\pi \tilde\epsilon^2}{\tau_2}} \,,
\end{equation}
where $\tilde\epsilon=\epsilon\tau_2 P_L/\sqrt{T_2 U_2-W_2^2/2}$.

This is to be supplemented by the path integral over the internal worldsheet fields. In particular, the contribution of the worldsheet fermion correlators after appropriate summation over the spin structures yields 
\begin{equation}
	\frac{1}{2}\sum_{a,b=0,1}(-1)^{a+b} \bigr(\langle \Psi(z)\bar\Psi(0)\rangle[^a_b]\bigr)^2 \vartheta[^a_b]^2\,\vartheta[^{a+H}_{b+G}]\,\vartheta[^{a-H}_{b-G}] = [\vartheta_1'(0)]^2 \,\vartheta[^{1+H}_{1+G}]\,\vartheta[^{1-H}_{1-G}]\,.
\end{equation}
This expression vanishes for $H=G=0$, reflecting  the $\mN=4$ subsector of the theory. The contribution of the twisted worldsheet fermions $\Psi^2, \Psi^3$ exactly cancels against that of their twisted bosonic superpartners $Z^2,Z^3$, as can be seen from the identity
\begin{equation}
	\Gamma_{4,4}[^H_G] = \frac{4\eta^6}{\vartheta[^{1+H}_{1+G}]^2}\,\left(\tfrac{1}{2}\sum_{\gamma,\delta=0,1}(-1)^{G(\gamma+H)}\,\bar\vartheta[^\gamma_\delta]^4-\tfrac{(-)^G}{2}\,\bar\vartheta[^{1+H}_{1+G}]^4\right) \,,
\end{equation}
valid for $(H,G)\neq(0,0)$. Furthermore, the contribution of $[\vartheta_1'(0)]^2=(2\pi\eta^3)^2$, together with the $\eta^6$ arising from the twisted lattice exactly cancels against the remaining $\eta^{-12}$ contribution of the left moving oscillators and the amplitude takes the manifestly BPS form
\begin{equation}
	\begin{split}
	\mc F(\epsilon) &= \int_{\mc F}\frac{d^2\tau}{\tau_2^2} \, \frac{G_{\rm bos}(\epsilon)}{4\bar\Delta}\sum_{\genfrac{}{}{0pt}{}{H,G=0,1}{(H,G)\neq(0,0)}} \sum_{\genfrac{}{}{0pt}{}{\gamma,\delta=0,1}{(\gamma,\delta)\neq (1,1)}}(-1)^{G(\gamma+1)}\,\bar\vartheta[^{\gamma+H}_{\delta+G}]^4 \,\Gamma_{2,18}[^H_G](T,U,W) \,,
	\end{split}
\end{equation}
where $\Delta(\tau)=\eta^{24}(\tau)$ is the modular discriminant and $\mathcal F$ is the fundamental domain of $\rm{SL}(2;\mb Z)$.

\subsubsection{Field theory limit}

We can now extract the field theory limit of the amplitude expanded around the ${\rm SU}(2)$ enhancement point $TU-W^2/2=-1$. Since states in the $H=1$ sector are characterised by non-trivial winding along the $X^1$ direction, the only relevant contribution stems from the sector $H=0, G=1$:
\begin{equation}
	\mc F(\epsilon)\bigr|_{\rm F.T.} =\tfrac{1}{2}\pi^2 \epsilon^2\sum_{m, n\atop Q=0}\int\frac{dt}{t} \,\frac{(-1)^{m_1}}{\sin^2 \epsilon\,t}\,e^{-t \sqrt{T_2 U_2-\frac{1}{2}W_2^2}\bar P_L} \,.
\end{equation}
Around the enhancement point, the amplitude exhibits a singularity due to the extra massless vectormultiplet states and includes the effect of the hypermultiplets charged under $\rm SU(2)$. For these contributions, the field theory limit of the amplitude takes the form
\begin{equation}
	\frac{1}{(2\pi\epsilon)^2}\,\mc F(\epsilon)\bigr|_{\rm F.T.} = \sum_{k=0,\pm 1}\left[\gamma_{\hbar}(k\mu)-\gamma_{\hbar}(k\mu+ m_{\rm h}) \right] \,,
\label{FieldTheorLimitNstar}
\end{equation}
where $\gamma_\epsilon(a)=\log \Gamma_2(a|\epsilon,-\epsilon)$ is given in terms of Barnes' double gamma function, and we have set $\hbar= 2\pi i \epsilon$. In addition, $\mu=1+ T  U- W^2/2$ and  $m_{\rm h}= U$ are the BPS mass parameters associated to the vector- and hyper- multiplets, respectively. As anticipated, our amplitude exactly reproduces the $\Omega$-deformed partition function of the $\mN=2^\star$ gauge theory \cite{Nekrasov:2002qd,Pestun:2007rz}.

%%%

%%%
The topological amplitudes $F_g$ evaluated at the one loop level in the heterotic string coupling, may be used to extract information about the dual type IIA compactification. More precisely, from the work of \cite{Bershadsky:1993cx,Bershadsky:1993ta}, these objects encode interesting topological information about the dual Calabi-Yau since they correspond, in the standard case, to the genus $g$ topological string free energy. The heterotic dilaton $S$ lies in a vector multiplet, which implies that the vector multiplet moduli space receives perturbative and non-perturbative corrections, whereas the hypermultiplet one is exact. In type IIA constructions based on Calabi-Yau manifolds, the dilaton lies in a hypermultiplet and the vector moduli space receives no corrections. A peculiarity of the realisation of $\Nstar$ in terms of the freely acting asymmetric orbifold of Section \ref{AsymOrb} is that, since all hypermultiplets have acquired Scherk-Schwarz masses and the corresponding moduli are stabilised at points $T_i,U_i\sim 1$, the would-be dual IIA dilaton on a Calabi-Yau is massive and stabilised at $g_{\rm II}\sim 1$. In other words, a perturbative type IIA description, dual to our asymmetric heterotic construction, does not exist. Nevertheless, the heterotic amplitudes $F_g$ are expected to probe topological information of a type II dual theory obtained as a Scherk-Schwarz reduction of M-theory.

\section{Geometrically engineering \texorpdfstring{$\mN=2^\star$  in String Theory}{N=2* realisation in String Theory}}\label{StringModel}

We have seen that the heterotic asymmetric orbifold  realisation of the previous section does not admit a perturbative type II dual. It is, however, possible to generalise our construction in order to geometrically engineer $\Nstar$ for more general orbifolds, including symmetric ones for which explicit dual type II descriptions are known.

\subsection{The general correspondence}

In order to construct a string theory uplift of the mass deformation in gauge theory, it is necessary to first elaborate on the properties of the latter since these are crucial in establishing our main correspondence. Indeed, as mentioned above, the mass deformation is responsible for the breaking of the $\mN=4$ supersymmetry to $\mN=2$ in such a way that the $\mN=2$ hypermultiplet inside the $\mN=4$ vectormultiplet acquires a mass. Clearly, the resulting theory -- the $\mN=2^\star$ gauge theory -- interpolates between the $\mN=4$ and the $\mN=2$ theories, in the zero and infinite mass limits, respectively.

The D-brane description turns out to be very useful in order to establish the correspondence. More precisely, the pure $\mN=4$ gauge theory can be realised as the theory of massless excitations of a stack of parallel D3-branes in which the degrees of freedom of the open strings stretching between the branes lead to the $\mN=4$ vector multiplet. In this picture, the mass deformation correponds to tilting the branes in some internal directions, in such a way that the mass be proportional to the angle. Recalling that branes at angles can alternatively be described by magnetic fluxes in the original background, one may, by analogy, expect that the mass deformation of gauge theory can also be described in string theory by a freely acting orbifold of the original $\mN=4$ compactification. It is very striking that this picture is quite generic and is due to the universality of the low-energy behaviour of a wide class of string models. Therefore, we propose a general correspondence for which we extensively argue below: \emph{freely acting orbifolds of $\mN=4$ compactifications realise  $\mN=2^\star$ in string theory.} More general constructions, such as flux compactifications corresponding to gaugings of $\mN=4$  supergravity are also expected to provide candidates for realising $\Nstar$ in string theory. However, they do not generically admit an exact worldsheet description.

We insist that, as explained below, the details of the specific construction of the freely-acting model do not impinge on the correspondence since the $\mN=2^\star$ features turn out ot be quite generic in the string moduli space, provided that its action preserves eight supercharges. This is what we refer to as \emph{universality}. In addition, the original $\mN=4$ theory is recovered in a standard fashion as a decompactification limit of the Scherk-Schwarz cycle.

In the present work, we mainly focus on compact internal spaces, in which case the gauge groups that can be described in the field theory are only subgroups of the stringy-allowed ones. Equivalently, anomaly cancellation restricts the number of allowed D-branes and, therefore, the maximum rank of the gauge group. However, one could consider non-compact internal spaces like $\mb C^2/\mb Z_k$ for arbitrary $k$ such that arbitrary ranks are permitted. Additionally, this leads to a natural way of creating a hierarchy of masses by taking the large $k$ limit while keeping the product of the Scherk-Schwarz radius $R$ with $k$ fixed, since the adjoint mass is then proportional to $1/kR$.

In what follows, we elaborate on the correspondence by analysing the properties of a broad class of freely-acting orbifolds. In particular, we show that the resulting spectrum is the $\mN=2^\star$ one.

\subsection{The class of models and their spectrum}\label{correspSubsec}

Without loss of generality, we focus on $\mb Z_2$ freely-acting orbifolds of $\mN=4$ heterotic compactifications for which we specify only the pertinent part of the orbifold action in order to keep the discussion generic, while deferring the discussion of specific models to subsequent sections. Starting from a $T^6$ compactification, the action of the orbifold should at least rotate the left-movers of two of the complex toroidal coordinates while shifting the third one, in such a way that the action does not have any fixed point. The partition function of this theory is

\begin{equation}
	Z=\frac{1}{2^2\,\eta^{12}\,\bar\eta^{24}}\sum_{\genfrac{}{}{0pt}{}{a,b=0,1}{H,G=0,1}}(-1)^{a+b+ab}\,\vartheta\bigr[^{a}_{b}\bigr]^2 \vartheta\bigr[^{a+H}_{b+G}\bigr]\vartheta\bigr[^{a-H}_{b-G}\bigr]\,\Gamma_{4,4}\bigr[^H_G\bigr] \,\Gamma_{2,18}\bigr[^H_G\bigr]\,,
\label{PartFnGen}
\end{equation}
where $a,b$ are summed over the spin structures on the worldsheet torus,  $H$ labels the orbifold sectors and the sum over $G$ imposes the $\mathbb Z_2$ invariant projection. $\Gamma_{4,4}\bigr[^H_G\bigr]$ is the twisted lattice partition function and $\Gamma_{2,18}\bigr[^H_G\bigr](T,U)$ is the shifted Narain lattice associated to the $T^2$ together with the gauge lattice directions. For convenience, we join the latter into a single lattice partition function $Z_{6,22}\bigr[^H_G\bigr]$. We further absorb potential phase factors depending on the specific model and ensuring modular invariance into the definition of $\Gamma_{2,18}$. It is clear that the orbifold action on the left movers is identical as that of the asymmetric model of Section \ref{AsymOrb} and preserves eight supercharges. Contrary to the asymmetric model, however, we now allow for a non-trivial action on the right-movers as well, incorporating also the case of symmetric actions on the $T^4$ coordinates.

The spectrum of the theory can be organised into contributions of the gravity, vector and hyper multiplets by expressing the partition function \eqref{PartFnGen} in terms of SO$(2n)$ characters:
\begin{equation}
	\begin{split}
	\eta^4\bar\eta^4\,Z =&(V_4 O_4-S_4 S_4)Z_{6,22}\bigr[^{\,0}_+\bigr]+(O_4 V_4-C_4 C_4)Z_{6,22}\bigr[^{\,0}_-\bigr]\\
	+&(V_4 C_4-S_4 V_4)Z_{6,22}\bigr[^{\,1}_\mp\bigr]+(O_4 S_4-C_4 O_4)Z_{6,22}\bigr[^{\,1}_\pm\bigr]\,.
	\end{split}
\label{PartFnGenChar}
\end{equation}
%\begin{equation}
%	\begin{split}
%	\eta^4\bar\eta^4\,Z =&(V_4 O_4-S_4 S_4)\bigr(\Gamma_{4,4}\bigr[^{\,0}_+\bigr]\Gamma_{2,18}\bigr[^{\,0}_+\bigr]+\Gamma_{4,4}\bigr[^{\,0}_-\bigr]\Gamma_{2,18}\bigr[^{\,0}_-\bigr]\bigr)\\
%	+&(O_4 V_4-C_4 C_4)\bigr(\Gamma_{4,4}\bigr[^{\,0}_-\bigr]\Gamma_{2,18}\bigr[^{\,0}_+\bigr]+\Gamma_{4,4}\bigr[^{\,0}_+\bigr]\Gamma_{2,18}\bigr[^{\,0}_-\bigr]\bigr)\\
%	+&(V_4 C_4-S_4 V_4)\bigr(\Gamma_{4,4}\bigr[^{\,1}_-\bigr]\Gamma_{2,18}\bigr[^{\,1}_+\bigr]+\Gamma_{4,4}\bigr[^{\,1}_+\bigr]\Gamma_{2,18}\bigr[^{\,1}_-\bigr]\bigr)\\
%	+&(O_4 S_4-C_4 O_4)\bigr(\Gamma_{4,4}\bigr[^{\,1}_+\bigr]\Gamma_{2,18}\bigr[^{\,1}_+\bigr]+\Gamma_{4,4}\bigr[^{\,1}_-\bigr]\Gamma_{2,18}\bigr[^{\,1}_-\bigr]\bigr)\,.
%	\end{split}
%\label{PartFnGenChar}
%\end{equation}
In the twisted sector $H=1$, the parity under the orbifold action of the lattice block $Z_{6,22}$ depends on whether the orbifold is asymmetric or symmetric. In addition, it can be straightforwardly expanded in terms of the toroidal and gauge lattices. For instance,
\begin{equation}
 Z_{6,22}\bigr[^{H}_+\bigr]=\Gamma_{4,4}\bigr[^{H}_+\bigr]\Gamma_{2,18}\bigr[^{H}_+\bigr]+\Gamma_{4,4}\bigr[^{H}_-\bigr]\Gamma_{2,18}\bigr[^{H}_-\bigr]\,,\textrm{ etc.}
\end{equation}
As in the asymmetric orbifold case, the  $H=1$ sector involves states with non-trivial winding and generically decouple in the low energy limit, provided the characteristic radius of the Scherk-Schwarz cycle is much larger than the string scale, $T_2/U_2\gg 1$. We hence focus our attention on the untwisted sector $H=0$.

Clearly, the states of interest around the SU(2) enhancement point $TU-W^2/2=-1$ are neutral with respect to the gauge bundle, $Q=0$, and arise only from the sectors:
\begin{equation}
	\begin{split}
				& (V_4 O_4-S_4 S_4)\,\Gamma_{4,4}\bigr[^{\,0}_+\bigr]\Gamma_{2,18}\bigr[^{\,0}_+\bigr] \\
				& (O_4 V_4-C_4 C_4) \,\Gamma_{4,4}\bigr[^{\,0}_+\bigr]\Gamma_{2,18}\bigr[^{\,0}_-\bigr] \,. 
	\end{split}
\end{equation}
Indeed, one obtains the vector multiplet  associated to the SU(2) gauge bosons, and the associated massive adjoint hypermultiplet. The quantum numbers, vertex operators, and masses of these states are precisely identical to those appearing in the asymmetric orbifold construction of Section \ref{AsymOrb}. This is a consequence of the fact that the SU(2) enhancement is realised by the Kaluza-Klein and winding states of the $T^2$ torus alone. In other words, they arise from a universal sector that is not affected by the gauge bundle, as can be seen by turning on generic Wilson lines.

Therefore, these models correctly engineer the $\mN=2^\star$ gauge theory in the field theory limit. This picture may be explicitly confirmed by calculating the partition function of the $\Omega$-deformation of the latter, as a field theory limit of the string theoretic graviphoton amplitude \eqref{ActionTopAmpl} in these general freely-acting orbifold backgrounds. Indeed, denoting by $\Phi[^H_G]$ the path integral over the internal worldsheet field (that depends on the specific details of the model), the amplitude takes form
\begin{equation}
	\begin{split}
	\mc F(\epsilon) &= \int_{\mc F}\frac{d^2\tau}{\tau_2^2} \,G_{\rm bos}(\epsilon)\sum_{\genfrac{}{}{0pt}{}{H,G=0,1}{}} \Phi[^H_G]\,\Gamma_{2,2}[^H_G] \,,
	\end{split}
\end{equation}
and one indeed recovers \eqref{FieldTheorLimitNstar}.

\subsection{An example heterotic/type II dual pair}

In order to illustrate our previous considerations, we consider an explicit example of heterotic/type II dual pair with $h^{1,1}=h^{2,1}=11$ discussed in \cite{Ferrara:1995yx,Gregori:1998fz}. We  briefly review their construction here. The starting point is the familiar heterotic/type II duality in six spacetime dimensions, with the type II theory living on ${\rm K3}\simeq T^4/\mathbb Z_2$ and the heterotic one on $T^4$. Upon toroidal compactification of each theory on a $T^2$, one obtains a four dimensional theory with $\mN=4$ supersymmetry. Subsequently, one introduces a freely acting $\mathbb Z_2'$ orbifold on both theories, which  is  responsible for the spontaneous breaking of $\mN=4\to \mN=2$. On the type IIA side, this free action generates an elliptic fibration of the Enriques surface over $\mathbb P^1\simeq T^2/\mathbb Z_2$. Explicitly, if we denote by $Z^1$ the complex coordinate of $T^2$, and $Z^2, Z^3$ the coordinates of $T^4$ in the type IIA theory, the orbifold action $g\in\mathbb Z_2$ generating the singular limit of K3 and the free action $G\in\mathbb Z_2' $ generating the elliptic fibration can be chosen to act as
\begin{equation}
	\begin{split}
		g: &\quad  Z^1\to Z^1 \quad,\quad Z^2\to -Z^2 \quad,\quad Z^3\to -Z^3 \,,\\
		G: &\quad Z^1\to -Z^1 \quad,\quad Z^2\to Z^2+i\pi\sqrt{\tfrac{T_2^{(2)}}{U_2^{(2)}} } \quad,\quad Z^3\to -Z^3+i\pi \sqrt{\tfrac{T_2^{(3)}}{U_2^{(3)}} }\,.
	\end{split}
\end{equation}
Here, $T^{(i)}$ and $U^{(i)}$ are the K\"ahler and complex structure moduli of the $T^2\times T^2\times T^2$ decomposition of the six dimensional internal space. By considering the six dimensional duality, it is straightforward to identify $T^{(1)}$ as the dual of  the heterotic dilaton modulus $S$. Similarly, the K\"ahler moduli of the K3 directions $T^{(2)}, T^{(3)}$ belonging to type IIA vector multiplets are mapped to the K\"ahler and complex structure moduli $T,U$ of the dual heterotic theory.

The action of $\mathbb Z_2'$ has no fixed points in the total internal space and, hence, there are no additional massless states arising from the twisted sectors. In addition to the $\mN=2$ gravity multiplet, one obtains an equal number $h^{1,1}=h^{2,1}=11$ of vector and hyper multiplets, as well as the tensor multiplet of the type IIA dilaton. The partition function of the theory explicitly reads
\begin{equation}
	\begin{split}
	Z_{\rm IIA} = \frac{1}{2^2\,\eta^{12}\bar\eta^{12}}\sum_{h,g=0,1\atop H,G=0,1}&\frac{1}{2}\sum_{a,b=0,1}(-1)^{a+b}\, \vartheta[^a_b]\,\vartheta[^{a+H}_{b+G}]\,\vartheta[^{a+h}_{b+g}]\,\vartheta[^{a-h-H}_{b-g-G}] \\
	\times &\frac{1}{2}\sum_{\bar a,\bar b=0,1}(-1)^{\bar a+\bar b+\bar a\bar b} \,\bar\vartheta[^{\bar a}_{\bar b}]\,\bar\vartheta[^{\bar a+H}_{\bar b+G}]\,\bar\vartheta[^{\bar a+h}_{\bar b+g}]\,\bar\vartheta[^{\bar a-h-H}_{\bar b-g-G}] \\
	\times & \ \Gamma_{2,2}^{(1)}\bigr[^{0}_{0}\bigr|^{H}_{G}\bigr] \ \Gamma_{2,2}^{(2)}\bigr[^{H}_{G}\bigr|^{h}_{g}\bigr]\  \Gamma_{2,2}^{(3)}\bigr[^{H}_{G}\bigr|^{h+H}_{g+G}\bigr] \,.
	\end{split}
\end{equation}
Here, $h,H=0,1$ label the orbifold sectors of $\mathbb Z_2\times \mathbb Z_2'$, and the sum over $g,G=0,1$ enforces the corresponding orbifold projections. The notation employed for the lattice sums associated to the three $T^2$ planes is as follows:
\begin{equation}
	\Gamma_{2,2}^{(j)}\bigr[^{h_1}_{g_1}\bigr|^{h_2}_{g_2}\bigr] = \left\{
				\begin{split}
							\frac{4\eta^3 \bar\eta^3}{\bigr|\vartheta\bigr[^{1+h_2}_{1+g_2}\bigr]\,\vartheta\bigr[^{1-h_2}_{1-g_2}\bigr]\bigr|} \quad &,\ {\rm for}\ (h_2,g_2)\neq(0,0)\ {\rm and}\ (h_1,g_1)\in\{(0,0),(h_2,g_2)\} \,,\\
							\Gamma_{2,2}^{(j),{\rm shift}}\bigr[^{h_1}_{g_1}\bigr]  ~~\quad\quad &, \ {\rm for}\ (h_2,g_2)=(0,0) \,,\\
							 0  \quad\qquad\qquad &, \ {\rm otherwise} \,.
				\end{split}
	\right.
\end{equation}
In the above notation for the shifted/twisted (2,2) lattice, the first column $[^{h_1}_{g_1}]$  denotes a shift of the two-torus along its first cycle, whereas the second column $[^{h_2}_{g_2}]$ denotes a twist. The vanishing components of the shifted/twisted lattice can be understood, for instance, from the fact that a lattice in the twisted sector with respect to the translation (shift) orbifold $(h_1,h_2)=(1,0)$ necessarily involves states with non-trivial momentum and/or winding numbers. On the other hand, a simultaneous projection with respect to the rotation (twist) orbifold $(g_1,g_2)=(0,1)$ only receives contributions from states with both momentum and winding numbers vanishing. By modular transformations, one shows the vanishing of the remaining sectors.

The dual theory may be constructed by consistently translating the free action of $\mathbb Z_2'$ on the heterotic degrees of freedom. It is clear that the action of $\mathbb Z_2'$ on the left movers must be identical to that of the freely acting asymmetric orbifold of Section \ref{AsymOrb}, and is responsible for the spontaneous breaking of $\mN=4\to \mN=2$. Since the type IIA model does have massless hyper multiplets, one is led to consider instead  a symmetric action of $\mathbb Z_2'$ on the $T^6$ internal space of the ${\rm E}_8\times {\rm E}_8$ heterotic string:
\begin{equation}
	\mathbb Z_2' : \quad Z^1 \to Z^1+i\pi \sqrt{\tfrac{T_2}{U_2}} \quad,\quad Z^2\to -Z^2 \quad,\quad Z^3\to -Z^3 \,.
\end{equation}
Determining the orbifold action on the heterotic gauge degrees of freedom is more subtle and requires a careful mapping of the $\mathbb Z_2'$ action on the 16 RR gauge fields arising from the twisted sector of the type IIA theory on ${\rm K3}\times T^2$. There, the shift action of  $\mathbb Z_2'$ has the effect of permuting the 16 fixed points, giving rise to 8 positive and 8 negative eigenvalues and reducing the rank of the ${\rm U}(1)^{16}$ gauge group by a factor of 2. On the heterotic side, this should be similarly mapped into an action on the ${\rm U}(1)^{16}$ Cartan generators of ${\rm E}_8\times {\rm E}_8$ with 8 positive and 8 negative eigenvalues.

Following \cite{Gregori:1998fz}, we consider a point in the moduli space where the ${\rm E}_8\times {\rm E_8}$ gauge group is broken to $\left({\rm SU}(2)\times {\rm SU}(2)\right)^8_{k=1} \simeq {\rm SO}(4)^8_{k=1}$. It is clear that a pairwise interchange of the two ${\rm SU}(2)$ gauge group factors introduces the correct number of negative eigenvalues, as required from the type IIA action of $\mathbb Z_2'$. Let us see how this is realised in the heterotic theory at the level of the partition function.

We first introduce three additional $\mathbb Z_2$ shift orbifolds, which correspond to discrete Wilson lines around the K3 directions. Each one of these is labeled by $h_i,g_i=0,1$ with $i=1,2,3$ and acts as a momentum shift along a separate $T^4$ direction, supplemented by an appropriate action on the gauge bundle degrees of freedom. It is most convenient to employ the fermionic realisation of ${\rm E_8\times E_8}$. The action of the $h_i,g_i$ is such that it breaks the gauge group down to ${\rm SO}(4)^8_{k=1}$ and its partition function contribution reads
\begin{equation}
	\begin{split}
	\Gamma_{8+8}\bigr[^{h_1,h_2,h_3}_{g_1,g_2,g_3}\bigr] = \tfrac{1}{2}\sum_{\gamma,\delta=0,1} &\bar\vartheta\bigr[^\gamma_\delta\bigr]^2\,\bar\vartheta\bigr[^{\gamma+h_1}_{\delta+g_1}\bigr]^2\,\bar\vartheta\bigr[^{\gamma+h_2}_{\delta+g_2}\bigr]^2\,\bar\vartheta\bigr[^{\gamma+h_3}_{\delta+g_3}\bigr]^2\\
	\times\	&\bar\vartheta\bigr[^{\gamma+h_1-h_2}_{\delta+g_1-h_2}\bigr]^2\,\bar\vartheta\bigr[^{\gamma+h_2-h_3}_{\delta+g_2-g_3}\bigr]^2\,\bar\vartheta\bigr[^{\gamma+h_3-h_1}_{\delta+g_3-g_1}\bigr]^2\,\bar\vartheta\bigr[^{\gamma-h_1-h_2-h_3}_{\delta-g_1-g_2-g_3}\bigr]^2  \,.
	\end{split}
\end{equation}
This describes a conformal system of 32 real fermions $\tilde\psi^a(\bar z)$ which are assembled into eight sets of four real fermions each. All fermions within a given set are assigned the same boundary conditions. Due to the different boundary conditions assigned to each of the eight sets, all Kac-Moody currents $\tilde J_{ab}(\bar z)=i\tilde\psi^a \tilde\psi^b$ are twisted except for those currents $\tilde J_{ab}$ formed out of real fermions belonging to the same set. Hence, each set of four real fermions realises an ${\rm SO}(4)_{k=1}$ gauge group, consistently with the fact that the central charge of  ${\rm SO}(4)_k$ is given by $c=6k/(k+2)$. The level one realisation of this current algebra  is obtained in terms of  four real fermions $\tilde\psi^1, \tilde\psi^2, \tilde\psi^3, \tilde\psi^4$ by forming the ${\rm SU}(2)_{+}\times {\rm SU}(2)_{-}$ currents
\begin{equation}
	\tilde J^a_{+}(\bar z) = \frac{i}{2}\left( \tfrac{1}{2}\,\epsilon^{abc}\tilde\psi^b\,\tilde\psi^c + \tilde\psi^a\,\tilde\psi^4\right)  \quad,\quad \tilde J^a_{-}(\bar z) = \frac{i}{2}\left( \tfrac{1}{2}\,\epsilon^{abc}\tilde\psi^b\,\tilde\psi^c - \tilde\psi^a\,\tilde\psi^4\right)  \,,
\end{equation}
where $a,b,c=1,2,3$. We now need to define the action of $\mathbb Z_2'$ on these fermionic variables, such that the two ${\rm SU}(2)_{k=1}$  factors are exchanged. It is clear that this can occur in two ways, either by twisting one of the four real fermions
\begin{equation}
	{\rm (i)} :\qquad \tilde\psi^a \to \tilde\psi^a \quad,\quad \tilde\psi^4 \to -\psi^4 \,,
\end{equation}
or by twisting three out of four
\begin{equation}
	{\rm (ii)} :\qquad	\tilde\psi^a \to -\tilde\psi^a \quad,\quad \tilde\psi^4 \to \psi^4 \,.
\end{equation}
Moreover, in order to generate 8 negative eigenvalues, as required by the action of $\mathbb Z_2'$ on the type IIA side, the twisting of fermions must be repeated for all eight sets, such that all eight  ${\rm SU}(2)\times {\rm SU}(2)$ factors are exchanged under it. This modifies the partition function contribution to
\begin{equation}
	\begin{split}
	\Gamma_{8+8}\bigr[^{h_i\,,\,H}_{g_i\,,\,G}\bigr] = \tfrac{1}{2}\sum_{\gamma,\delta=0,1}\  &\bar\vartheta\bigr[^{\gamma+H}_{\delta+G}\bigr]^{3/2}\,\bar\vartheta\bigr[^{\gamma+H+h_1}_{\delta+G+g_1}\bigr]^{3/2}\,\bar\vartheta\bigr[^{\gamma+h_2}_{\delta+g_2}\bigr]^{3/2}\,\bar\vartheta\bigr[^{\gamma+h_3}_{\delta+g_3}\bigr]^{3/2} \,(-1)^{G(\gamma+H)}\\
	\times\	&\bar\vartheta\bigr[^{\gamma+h_1-h_2}_{\delta+g_1-g_2}\bigr]^{3/2}\,\bar\vartheta\bigr[^{\gamma+h_2-h_3}_{\delta+g_2-g_3}\bigr]^{3/2}\,\bar\vartheta\bigr[^{\gamma+h_3-h_1}_{\delta+g_3-g_1}\bigr]^{3/2}\,\bar\vartheta\bigr[^{\gamma-h_1-h_2-h_3}_{\delta-g_1-g_2-g_3}\bigr]^{3/2}  \\
	\times\  & \bar\vartheta\bigr[^\gamma_\delta\bigr]^{1/2}\,\bar\vartheta\bigr[^{\gamma+h_1}_{\delta+g_1}\bigr]^{1/2} 
		\bar\vartheta\bigr[^{\gamma+H+h_2}_{\delta+G+g_2}\bigr]^{1/2}\,\bar\vartheta\bigr[^{\gamma+H+h_3}_{\delta+G+g_3}\bigr]^{1/2}\\
	\times\	&\bar\vartheta\bigr[^{\gamma+H+h_1-h_2}_{\delta+G+g_1-g_2}\bigr]^{1/2}\,\bar\vartheta\bigr[^{\gamma+H+h_2-h_3}_{\delta+G+g_2-g_3}\bigr]^{1/2}\,\bar\vartheta\bigr[^{\gamma+H+h_3-h_1}_{\delta+G+g_3-g_1}\bigr]^{1/2}\,\bar\vartheta\bigr[^{\gamma+H-h_1-h_2-h_3}_{\delta+G-g_1-g_2-g_3}\bigr]^{1/2}  \,.
	\end{split}
	\label{LatticeTwistSU2x8}
\end{equation}
For two of the sets, we have chosen to twist three real fermions, whereas for the remaining six sets we twisted one real fermion. The above partition function displays the reduction of the rank of the gauge group from 16 to 8 due to the $\mathbb Z_2'$ action. Indeed, only the diagonal ${\rm SU}(2)^8$ remains after the twisting, and each ${\rm SU}(2)_{k=2}$ factor can be realised  at level two by three real fermions, $\tilde J^{ab}=i\tilde\psi^a \tilde\psi^b$, with $a,b=1,2,3$.

Of course, in order to compare with the perturbative spectrum of the type IIA construction, the heterotic ${\rm SU}(2)^8$ gauge group must be broken down to its abelian subgroup ${\rm U}(1)^{8}$. This is achieved by introducing an additional discrete Wilson line, parametrised by $h_4,g_4=0,1$ acting as a momentum shift along the fourth direction of $T^4$ and responsible for twisting one of the real fermions that realise each ${\rm SU}(2)$. Simultaneously, it further twists the eight real fermions corresponding to the $\bar\vartheta^{1/2}$ of the last two lines of \eqref{LatticeTwistSU2x8}. One then obtains the following contribution to the gauge bundle partition function
\begin{equation}
	\begin{split}
	\Gamma_{8+8}\bigr[^{h_i\,,\,H}_{g_i\,,\,G}\bigr] = &\, \tfrac{1}{2}\sum_{\gamma,\delta=0,1} \bar\vartheta\bigr[^{\gamma-H}_{\delta-G}\bigr]\,\bar\vartheta\bigr[^{\gamma+H+h_1}_{\delta+G+g_1}\bigr]\,\bar\vartheta\bigr[^{\gamma+h_2}_{\delta+g_2}\bigr]\,\bar\vartheta\bigr[^{\gamma+h_3}_{\delta+g_3}\bigr] \\
	\times\	&\bar\vartheta\bigr[^{\gamma+h_1-h_2}_{\delta+g_1-g_2}\bigr]\,\bar\vartheta\bigr[^{\gamma+h_2-h_3}_{\delta+g_2-g_3}\bigr]\,\bar\vartheta\bigr[^{\gamma+h_3-h_1}_{\delta+g_3-g_1}\bigr]\,\bar\vartheta\bigr[^{\gamma-h_1-h_2-h_3}_{\delta-g_1-g_2-g_3}\bigr]  \\
	\times\  & \bar\vartheta\bigr[^{\gamma+h_4}_{\delta+g_4}\bigr]^{1/2}\,\bar\vartheta\bigr[^{\gamma+h_4+h_1}_{\delta+g_4+g_1}\bigr]^{1/2} 
\bar\vartheta\bigr[^{\gamma+h_4+h_2}_{\delta+g_4+g_2}\bigr]^{1/2}\,\bar\vartheta\bigr[^{\gamma+h_4+h_3}_{\delta+g_4+g_3}\bigr]^{1/2} \\
	\times\	&\bar\vartheta\bigr[^{\gamma+h_4+h_1-h_2}_{\delta+g_4+g_1-g_2}\bigr]^{1/2}\,\bar\vartheta\bigr[^{\gamma+h_4+h_2-h_3}_{\delta+g_4+g_2-g_3}\bigr]^{1/2}\,\bar\vartheta\bigr[^{\gamma+h_4+h_3-h_1}_{\delta+g_4+g_3-g_1}\bigr]^{1/2}\,\bar\vartheta\bigr[^{\gamma+h_4-h_1-h_2-h_3}_{\delta+g_4-g_1-g_2-g_3}\bigr]^{1/2}\\		
	\times\		&\bar\vartheta\bigr[^{\gamma+h_4+H}_{\delta+g_4+G}\bigr]^{1/2}\,\bar\vartheta\bigr[^{\gamma+h_4+H+h_1}_{\delta+g_4+G+g_1}\bigr]^{1/2}\,\bar\vartheta\bigr[^{\gamma+h_4+H+h_2}_{\delta+g_4+G+g_2}\bigr]^{1/2}\,\bar\vartheta\bigr[^{\gamma+h_4+H+h_3}_{\delta+g_4+G+g_3}\bigr]^{1/2}\\
	\times\	&\bar\vartheta\bigr[^{\gamma+h_4+H+h_1-h_2}_{\delta+g_4+G+g_1-g_2}\bigr]^{1/2}\,\bar\vartheta\bigr[^{\gamma+h_4+H+h_2-h_3}_{\delta+g_4+G+g_2-g_3}\bigr]^{1/2}\,\bar\vartheta\bigr[^{\gamma+h_4+H+h_3-h_1}_{\delta+g_4+G+g_3-g_1}\bigr]^{1/2}\,\bar\vartheta\bigr[^{\gamma+h_4+H-h_1-h_2-h_3}_{\delta+g_4+G-g_1-g_2-g_3}\bigr]^{1/2}  \,.
		\end{split}
\end{equation}
The total partition function of the heterotic dual theory then reads
\begin{equation}
	\begin{split}
		Z_{\rm het} = \frac{1}{2^5\,\eta^{12}\,\bar\eta^{24}}\sum_{H,G=0,1 \atop h_i,g_i=0,1} \frac{1}{2}\sum_{a,b=0,1}(-1)^{a+b+ab}\,\vartheta\bigr[^a_b\bigr]^2\,\vartheta\bigr[^{a+H}_{b+G}\bigr]\,\vartheta\bigr[^{a-H}_{b-G}\bigr]\,\Gamma_{2,2}\bigr[^{H\,,\,h_1}_{G\,,\,g_1}\bigr]\,\Gamma_{4,4}\bigr[^{h_i}_{g_i}\bigr|^H_G\bigr]\,\Gamma_{8+8}\bigr[^{h_i\,,\,H}_{g_i\,,\,G}\bigr]\,,
	\end{split}
\end{equation}
where $(H,G)$ acts on the (2,2) lattice as a momentum shift along the first cycle of $T^2$,  $(h_1,g_1)$ acts on the (2,2) lattice as a winding shift along the same cycle, and the four $(h_i,g_i)$ orbifolds act on the (4,4) lattice as momentum shifts along four distinct cycles of $T^4$.  Because of the simultaneous twist by $(H,G)$ and the shifts by $(h_i,g_i)$ along the four directions of $T^4$, the $\mN=2$ subsectors $(H,G)\neq(0,0)$ only receive contributions from $(h_i,g_i)=(0,0)$ or $(h_i,g_i)=(H,G)$ for each $i=1,2,3,4$. Notice that a momentum shift due to $(H,G)$ together with a winding shift due to $(h_1,g_1)$ along the same direction of $T^2$ introduces a level matching asymmetry in the (2,2) lattice which, in turn, ensures that the level matching constraints of the entire theory are properly satisfied and the  partition function $Z_{\rm het}$ is modular invariant.

Using the explicit form of their respective partition functions, it is straightforward to show that the massless spectra of the type IIA and heterotic models indeed match. Furthermore, it is easy to show that the vector multiplets arising from the $T^2$ gauge group, including the ${\rm SU}(2)$ enhancement at $TU=-1$, together with the massive hypermultiplets  charged under it, survive the projections and the form of their corresponding vertex operators is identical to those in the asymmetric model of Section \ref{AsymOrb}. The fact that this orbifold construction  correctly reproduces  \eqref{FieldTheorLimitNstar} in the field theory limit and, hence, geometrically engineers $\Nstar$ gauge theory, is a natural consequence of the arguments given in Section \ref{correspSubsec}.

The recovery of $\mN=4$ supersymmetry occurs in the decompactification limit of $T^2$ where the Scherk-Schwarz radius is taken to infinity, $R_1\to\infty$,  whereas in the limit $R_1\to 0$, one recovers a standard $\mN=2$ theory as in a non-freely acting orbifold. This property is precisely reproduced in the type IIA dual, by considering the large and small volume limits in the $T^{(2)}, T^{(3)}$ moduli. This implies that the map between the heterotic and type IIA moduli is, in fact, linear and the correspondence can be extracted explicitly \cite{Gregori:1998fz} by a careful analysis of gravitational threshold corrections on both sides, yielding $S=T^{(1)}/8\pi$, $T=T^{(2)}$ and $U=T^{(3)}$.

To further our check of this realisation of $\Nstar$, we again compute the topological amplitudes $F_g$ of eq. \eqref{ActionTopAmpl} at genus one in heterotic perturbation theory:
\begin{equation}
	\begin{split}
	\mc F(\epsilon) &= \int_{\mc F}\frac{d^2\tau}{\tau_2^2} \, \frac{G_{\rm bos}(\epsilon)}{\bar\Delta}\sum_{\genfrac{}{}{0pt}{}{H,G=0,1}{(H,G)\neq(0,0)}} \,\left( \Gamma_{2,2}^{\lambda=0}\bigr[^H_G\bigr]\,\bar\Phi^{(0)}\bigr[^H_G\bigr]  + \Gamma_{2,2}^{\lambda=1}\bigr[^H_G\bigr]\,\bar\Phi^{(1)}\bigr[^H_G\bigr]    \right)\,,
	\end{split}
\end{equation}
where the forms $\Phi^{(\lambda)}$ are given by
\begin{align}\nonumber
	\Phi^{(0)}\bigr[^0_1\bigr] &=\tfrac{1}{2}\,(\vartheta_3^4+\vartheta_4^4)\,\vartheta_3^8\,\vartheta_4^8 \,, \\ 
	\Phi^{(0)}\bigr[^1_0\bigr] &=-\tfrac{1}{2}\,(\vartheta_2^4+\vartheta_3^4)\,\vartheta_2^8\,\vartheta_3^8 \,, \\ \nonumber
	\Phi^{(0)}\bigr[^1_1\bigr] &=\tfrac{1}{2}\,(\vartheta_2^4-\vartheta_4^4)\,\vartheta_2^8\,\vartheta_4^8 \,,\\ 
\intertext{and} \nonumber
	\Phi^{(1)}\bigr[^0_1\bigr] &=\vartheta_3^{10}\,\vartheta_4^{10} \,, \\ 
	\Phi^{(1)}\bigr[^1_0\bigr] &=-\vartheta_2^{10}\,\vartheta_3^{10} \,, \\ \nonumber
	\Phi^{(1)}\bigr[^1_1\bigr] &=-\vartheta_2^{10}\,\vartheta_4^{10} \,. 
\end{align}
The notation $\Gamma_{2,2}^{\lambda}[^H_G]$ for the $T^2$ lattice denotes a momentum shift along the first cycle for $\lambda=0$, whereas for $\lambda=1$ it stands for a simultaneous momentum and winding shift along the same cycle. The contribution involving the $\lambda=1$ lattice decouples in the field theory limit since  $\Gamma^{\lambda=1}$ involves states which simultaneously carry non-trivial momentum and winding numbers. Their masses are of the form $\sim 1/R^2+R^2$ and are therefore of the order of the string scale. Consequently, only the $\lambda=0$ sector leads to a non trivial field theory limit. Repeating the analysis of Section \ref{AsymOrb}, one similarly recovers the correct gauge theory partition function \eqref{FieldTheorLimitNstar} at the $SU(2)$ symmetry enhancement point.

%%%%%%%%%%%%%%%%%%%%%%%%%%%%%

\section{Type I description of the mass deformation}\label{TypeOne}

So far, we have presented a heterotic realisation of $\Nstar$ in terms of a freely-acting asymmetric orbifold in Section \ref{AsymOrb} and its symmetric generalisations in Section \ref{StringModel}, the latter naturally admitting dual type II descriptions on K3 fibered Calabi-Yau manifolds. In this section, we discuss the analogous realisation of $\Nstar$ by performing an orientifold of the type IIB theory on a freely acting version of ${\rm K3}\times T^2$ implementing the Scherk Schwarz breaking $\mN=4\to\mN=2$.

In the heterotic construction of Section \ref{AsymOrb}, the main ingredients were: {\it (i)} the spontaneous nature of the breaking of $\mN=4\to\mN=2$ via a Scherk-Schwarz flux, {\it (ii)} the gauge group being unaffected by the orbifold action, and {\it (iii)} rendering adjoint hypermultiplet states massive. To this end, the simplest prototype type I vacuum realising $\Nstar$ may be naturally identified with a freely acting $\mathbb Z_2$ orbifold of the standard $\mN=4$ type I theory on $T^6$ with ${\rm SO}(32)$ gauge group. This model was first constructed in \cite{Antoniadis:1998ep} in the context of partial supersymmetry breaking in type I theory.

The $\mathbb Z_2$ orbifold acts as a momentum shift along one of the complexified string coordinates, supplemented by a reflection of the remaining two:
\begin{equation}\label{shifttwistaction}
	Z^1 \to Z^1 + i\pi \sqrt{\tfrac{T_2}{U_2}} \quad,\quad Z^2 \to - Z^2 \quad,\quad Z^3\to - Z^3 \,,
\end{equation}
and the torus partition function reads
\begin{equation}
	\begin{split}
	\mathcal T= \frac{1}{2\,\eta^{12}\,\bar\eta^{12}}\sum_{H,G=0,1} &\frac{1}{2}\sum_{a,b} (-1)^{a+b+ab}\,\vartheta\bigr[^a_b\bigr]^2\,\vartheta\bigr[^{a+H}_{b+G}\bigr]\,\vartheta\bigr[^{a-H}_{b-G}\bigr]\  \\
	\times & \frac{1}{2}\sum_{\bar a,\bar b} (-1)^{\bar a+\bar b+\bar a\bar b}\,\bar\vartheta\bigr[^{\bar a}_{\bar b}\bigr]^2\,\bar\vartheta\bigr[^{\bar a+H}_{\bar b+G}\bigr]\,\bar\vartheta\bigr[^{\bar a-H}_{\bar b-G}\bigr]\ \\
	\times & \Gamma_{2,2}\bigr[^H_G\bigr](T,U) \ \Gamma_{4,4}\bigr[^H_G\bigr] \,,
	\end{split}
\end{equation}
where the (2,2) lattice is shifted, while the (4,4) one is twisted as indicated in \eqref{shifttwistaction}. Clearly, the perturbative closed string sector enjoys unbroken $\mathcal N=(2,2)$ supersymmetry and, hence, cannot give rise to non-abelian gauge symmetry. It is irrelevant for the purposes of this discussion, and we henceforth focus entirely on the open string sector. The relevant diagrams here are the annulus and the M\"obius strip.

Following \cite{Antoniadis:1998ep}, one may construct a consistent vacuum with $N=32$ D9 branes, where the  $\mathbb Z_2$ orbifold acts trivially on the Chan-Paton degrees of freedom. The annulus and M\"obius strip amplitudes in the direct channel read
\begin{equation}\label{Annul}
	\mathcal A =  \frac{N^2}{4\,\eta^{12}} \sum_{G=0,1} \frac{1}{2}\sum_{a,b} (-1)^{a+b+ab}\,\vartheta\bigr[^a_b\bigr]^2\,\vartheta\bigr[^{\ \,a}_{b+G}\bigr]\,\vartheta\bigr[^{\ \,a}_{b-G}\bigr]\  \Gamma_{2}^{(1)}\bigr[^{\,0}_G\bigr] \,\Gamma_{4}\bigr[^{\,1}_G\bigr] \,,
\end{equation}

\begin{equation}\label{Moebi}
	\mathcal M = -\frac{N}{4\,\hat\eta^{12}} \sum_{G=0,1} \frac{1}{2}\sum_{a,b} (-1)^{a+b+ab}\,\hat\vartheta\bigr[^a_b\bigr]^2\,\hat\vartheta\bigr[^{\ \,a}_{b+G}\bigr]\,\hat\vartheta\bigr[^{\ \,a}_{b-G}\bigr]\  \Gamma_{2}^{(1)}\bigr[^{\,0}_G\bigr] \,\hat\Gamma_{4}\bigr[^{\,1}_G\bigr] \,,
\end{equation}
with $\Gamma_2$ being the shifted $T^2$ lattice partition function
\begin{equation}
	\Gamma_2^{(1)}\bigr[^{\,0}_G\bigr]  =\sum_{m_1,m_2\in \mathbb Z} (-1)^{ m_1 G}\,e^{-\pi t|P|^2} \,,
\end{equation}
defined in terms of the real Schwinger parameter $t$ and the lattice momenta
\begin{equation}
	P = \frac{ m_2-U m_1 }{\sqrt{T_2 U_2} }\,.
\end{equation}
Furthermore, $\Gamma_4$ is the twisted $T^4$ partition function
\begin{equation}
	\Gamma_4 \bigr[^{\, 0}_G\bigr] = \left\{ \begin{split}
												\Gamma_2^{(2)}\bigr[^0_0\bigr] \,\Gamma_2^{(3)}\bigr[^0_0\bigr]  \quad,\quad G=0\\
												\left(\frac{2\eta^3}{\vartheta_2}\right)^2 \quad\quad,\quad G=1
								\end{split} \right.  \,.
\end{equation}
The arguments of the Dedekind $\eta(\tau)$ and Jacobi $\vartheta\bigr(\tau)$ functions in the annulus and M\"obius amplitudes above are understood to be $\tau=\frac{it}{2}$ and $\tau=\frac{it}{2}+\frac{1}{2}$, respectively, and the hatted characters denote the real basis for the M\"obius amplitude \cite{Bianchi:1990yu,Bianchi:1990tb}. The Chan-Paton factors $N^2$ and $N$ simply reflect the fact that each of the string endpoints is attached to one of the $N$ D9's in the case of the annulus, or to the fact that the string is stretched between one of the $N$ D9's and an O9 plane, in the case of the M\"obius.

To extract the low lying spectrum, it is convenient to decompose the partition functions in terms of characters:
\begin{equation}
	\mathcal A=\frac{1}{4\eta^8}\,N^2 \,\sum_{m_1,m_2\in\mathbb Z}\left[ (Q_o+Q_v)\,\Gamma^{(1)}_{m_1,m_2}\,\Gamma_2^{(2)}\,\Gamma_2^{(3)}+(Q_o-Q_v)\,\left(\frac{2\eta^3}{\theta_2}\right)^2\,(-1)^{m_1}\,\Gamma^{(1)}_{m_1,m_2}\right] \,,
\end{equation}

\begin{equation}
	\mathcal M=-\frac{1}{4\hat\eta^8}\,N \,\sum_{m_1,m_2\in\mathbb Z}\left[ (\hat Q_o+\hat Q_v)\,\Gamma^{(1)}_{m_1,m_2}\,\Gamma^{(2)}\,\Gamma^{(3)}+(\hat Q_o-\hat Q_v)\,\left(\frac{2\hat\eta^3}{\hat\theta_2}\right)^2\,(-1)^{m_1}\,\Gamma^{(1)}_{m_1,m_2}\right] \,.
\end{equation}
We employ here the traditional ${\rm SO}(2n)$ character notation $Q_o = V_4 O_4 -S_4 S_4$ and $Q_v = O_4 V_4-C_4 C_4$, corresponding to the $\mN =2$ vector and hyper multiplets, respectively. Furthermore, we have set $\Gamma_{m_1,m_2}^{(1)} = e^{-\pi t |P|^2}$. 

It is straightforward to see that the theory has ${\rm SO}(32)$ gauge symmetry with all hypers transforming in the adjoint representation of ${\rm SO}(32)$, and carrying non-trivial Scherk-Schwarz mass. For lowest lying such states, the mass reads $m^2 = |U|^2/T_2 U_2$.  Moreover, by Poisson resumming the $T^2$ lattice momenta of $\Gamma_2^{(1)}$, it is easy to see that the limit $T_2 \to \infty$ in  \eqref{Annul} and \eqref{Moebi} projects onto the $G=0$ sector and, hence, one recovers $\mN=4$ supersymmetry. In fact, the type I vacuum discussed above is in several ways similar to the asymmetric orbifold construction of Section \ref{AsymOrb} and possesses all the properties that one expects from a string theoretic realisation of the $\Nstar$ gauge theory.

Before we discuss the calculation of topological amplitudes in this background, it is necessary to consider first a deformation of this theory in which the ${\rm SO}(32)$ gauge symmetry is broken to the Coulomb branch ${\rm U}(1)^{16}$ by turning on generic Wilson lines around the $T^2$. This deformation then acts as a natural parameter controlling the enhancement when we consider the topological amplitudes.  A standard parametrisation of the Wilson line background for each boundary is
\begin{equation}
	U = \bigoplus_{i=1}^{16} \, e^{2\pi i W^i \sigma_3}\,,
\end{equation}
with $\sigma_3$ being the Pauli matrix and $W_i$ being the Wilson lines in the Cartan sub-algebra of the ${\rm SO}(32)$ gauge group. Taking into account that, in the transverse channel, the Chan-Paton factors for the annulus and M\"obius become $({\rm Tr}\, U)^2$ and ${\rm Tr}\, (U^2)$, respectively, one finds in the direct channel
\begin{equation}
	\mathcal A =  \frac{1}{4\,\eta^{12}} \sum_{G=0,1} \frac{1}{2}\sum_{a,b} (-1)^{a+b+ab}\,\vartheta\bigr[^a_b\bigr]^2\,\vartheta\bigr[^{\ \,a}_{b+G}\bigr]\,\vartheta\bigr[^{\ \,a}_{b-G}\bigr]\  \tilde\Gamma_{2}^{(1)}\bigr[^{\,0}_G\bigr] \,\Gamma_{4}\bigr[^{\,1}_G\bigr] \,.
\end{equation}
For later convenience, the Chan-Paton factors have been entirely absorbed into the modified lattice sum
\begin{equation}
	\tilde \Gamma_2^{(1)} \bigr[^{\, 0}_G\bigr] =\sum_{m_1,m_2\in \mathbb Z} \ \sum_{i,j=1}^{16}\ \sum_{q_1,q_2=\pm 1}(-1)^{m_1 G}\exp\left[-\pi t\,\frac{|m_2-Um_1+q_1 W^i+q_2 W^j |^2}{T_2 U_2-\frac{1}{2}W_2^2}\right] \,.
\end{equation}
Similarly, for the M\"obius strip one finds
\begin{equation}
	\mathcal M = -\frac{1}{4\,\hat\eta^{12}} \sum_{G=0,1} \frac{1}{2}\sum_{a,b} (-1)^{a+b+ab}\,\hat\vartheta\bigr[^a_b\bigr]^2\,\hat\vartheta\bigr[^{\ \,a}_{b+G}\bigr]\,\hat\vartheta\bigr[^{\ \,a}_{b-G}\bigr]\  \check\Gamma_{2}^{(1)}\bigr[^{\,0}_G\bigr] \,\hat\Gamma_{4}\bigr[^{\,1}_G\bigr] \,,
\end{equation}
with 
\begin{equation}
	\check\Gamma_{2}^{(1)}\bigr[^{\,0}_G\bigr] =\sum_{m_1,m_2\in \mathbb Z} \ \sum_{i=1}^{16}\ \sum_{q_1=\pm 1}(-1)^{m_1 G}\exp\left[-\pi t\,\frac{|m_2-Um_1+2q_1 W^i |^2}{T_2 U_2-\frac{1}{2}W_2^2}\right] \,.
\end{equation}

The spectrum of the theory contains 16 massless vector multiplets, corresponding to the Cartan sub-algebra ${\rm U}(1)^{16}$. The charged vectors, corresponding to the roots $\rho$ of ${\rm SO}(32)$ are instead massive. For the low lying states without Kaluza-Klein momenta, the mass reads
\begin{equation}
	m_\rho^2 = \frac{ |\rho\cdot W|^2}{T_2 U_2-\frac{1}{2}W_2^2} \,.
\end{equation}
The hyper multiplets, on the other hand, are  massive, as they carry odd momentum number $m_1 \in 2\mathbb Z+1$, and the low lying states have
\begin{equation}\label{hypmassform}
	m_Q^2 = \frac{ | U+Q\cdot W |^2 }{T_2 U_2-\frac{1}{2}W_2^2} \,.
\end{equation}
Here, $Q^i$ is the charge vector with respect to the Cartan subalgebra of ${\rm SO}(32)$ associated to the adjoint representation. For simplicity, we consider an ${\rm SU}(2)$ enhancement point obtained by choosing $\mu \equiv W_1 -W_2\to 0$. Then, clearly, the charged vectors with $Q=Q_\pm\equiv \pm(1,-1,0,\ldots,0)$ become massless and enhance to an ${\rm SU}(2)$. Similarly, the hyper multiplets transforming in the adjoint of ${\rm SU}(2)$ carry masses given by \eqref{hypmassform} for $Q=Q_\pm$ and $Q=(0,\ldots,0)$.

We would like to confirm the result above by calculating the topological amplitude $F_g$ perturbatively in this theory. In order to do so, we recall the vertex operator of the graviphoton in the orientifold theory:
\begin{align}
V(p,\epsilon)=\epsilon_\mu &\bigg[\left(\partial Z^1+i(p\cdot \Psi) \Psi^1\right) \left(\bar{\partial} Z^{\mu}+i(p\cdot\tilde{\Psi}) 
\tilde{\Psi}^\mu\right)+(\rm{left} \leftrightarrow \rm{right})\nonumber\\
-&e^{-\tfrac{1}{2}(\varphi+\tilde{\varphi})} p_\nu S^\alpha {(\sigma^{\mu\nu})_\alpha}^\beta \tilde{S}_\beta\,e^{\frac{i}{2}(\Phi_1+\tilde{\Phi}_1)}\, 
\epsilon^{AB}\Sigma_A \tilde{\Sigma}_B\bigg] \,e^{ip\cdot Z}\,.\label{TypeIVertexGTtype}
\end{align}
Here, $\epsilon$ and $p$ are the polarisation and momentum vectors of the operator, respectively, satisfying $\epsilon\cdot p=0$, and $\varphi$ is the ghost field. In addition, the tilded fields are the right-moving degrees of freedom. 
Since \eqref{TypeIVertexGTtype} leads solely to deformations of the worldsheet sigma model in the space-time directions, the calculation goes along the same lines as in \cite{Gava:1996hr,AFHNZ}. More precisely, the path integral over the space-time directions can be obtained through the generating function
\begin{align}
G_{\text{bos}}(\epsilon)=\left\langle \exp\Biggr[\frac{\hat\epsilon}{t}\int d^2\sigma
\left(Z^4(\bar\partial-\partial) Z^5+ \bar Z^4(\bar\partial-\partial)\bar Z^5\right)\Biggr]\right\rangle ~,\label{TypeIBosSpaceTime}
\end{align}
for the bosonic fields and
\begin{align}\label{SpacetimeFermDef}
 G_{\text{ferm}}(\epsilon)= \left\langle \exp\Biggr[\frac{\hat\epsilon}{t}\int 
d^2\sigma \left[ (\Psi^4-\tilde\Psi^4)(\Psi^5-\tilde\Psi^5)+(\bar\Psi^4-\tilde{\bar{\Psi}}^4)(\bar\Psi^5-\tilde{\bar{\Psi}}^5)
\right] \Biggr]\right\rangle~,
\end{align}
for the fermionic ones. Here, we have introduced the notation $\hat\epsilon=\epsilon P t/\sqrt{T_2 U_2-W_2^2/2}$. As argued above, we are only interested in the contributions of the annulus and the M\"obius strip. To this end, the mode expansions for the worldsheet fields are
\begin{align}\label{ModesZ}
& Z^i = \sum\limits_{n,m} Z^i_{n,m}\cos(\pi n\sigma_2)e^{2\pi i m \sigma_1}\,,&& \bar Z^i = \sum\limits_{n,m} 
\bar{Z}^i_{n,m}\cos(\pi n\sigma_2)e^{2\pi i m \sigma_1}\,,\\
&\Psi^i = \sum\limits_{n,m}\Psi^i_{n,m}\, e^{\pi i(n\sigma_2+2m \sigma_1)}~,  &&\tilde\Psi^i = \sum\limits_{n,m}\Psi^i_{n,m}\, 
e^{\pi i(-n\sigma_2+2m\sigma_1)}\,,\\
&\bar\Psi^i = \sum\limits_{n,m}\bar\Psi^i_{n,m}\, e^{\pi i (n\sigma_2+2m \sigma_1)}~,  &&\tilde{\bar{\Psi}}^i = 
\sum\limits_{n,m}\bar\Psi^i_{n,m}\, e^{\pi i(-n\sigma_2+2m \sigma_1)}\,.
\end{align}
for the annulus, whereas for the M\"obius strip we use the same expansion with the constraint $m+n\in2\mb Z$ and replace $t$ by $2t$. As a consequence of the BPS property, after summing over spin structures, the fermionic path integral yields a vanishing contribution for $G=0$ (as expected from an $\mN=4$ sector), whereas for $G=1$ the bosonic and fermionic path integrals for both the annulus and M\"obius worldsheets cancel each other except for the zero modes $n=0$ for which only the bosonic fields contribute:
\begin{equation}
G^{\mc A,\mc M}(\epsilon)=\frac{\pi^2\hat\epsilon^2}{\sin^2(\pi\hat\epsilon)}\,.
\end{equation}
Notice that, despite the different boundary conditions between the annulus and the M\"obius strip, both correlators are equal. The reason is that for the $n=0$ modes, the fact that $m$ is even for the strip is compensated by the doubling of the Teichm\"uller  parameter. Finally, the path integral over the internal degrees of freedom is trivial and, hence, the total contribution to $F_g$ is
\begin{equation}
\mc F^{\mc A+\mc M}(\epsilon)=\frac{1}{4}\int \frac{dt}{t}G^{\mc A}(\epsilon)\left(\tilde\Gamma_{2}^{(1)}\bigr[^{0}_1\bigr]-\check\Gamma_{2}^{(1)}\bigr[^{0}_1\bigr]\right)\,.
\label{topAmplTypeI}
\end{equation}
Note, however, that this lattice combination yields precisely the BPS mass formula
\begin{equation}
	\frac{1}{2}\left(\tilde\Gamma_{2}^{(1)}\bigr[^{0}_1\bigr]-\check\Gamma_{2}^{(1)}\bigr[^{0}_1\bigr]\right) = \sum_{m_1,m_2\in \mathbb Z} \ \sum_{Q\in {\rm Adj} } (-1)^{m_1}\exp\left[-\pi t\,\frac{|m_2-Um_1+Q\cdot W |^2}{T_2 U_2-\frac{1}{2}W_2^2}\right] \,,
\end{equation}
with the Cartan charges $Q$ being now summed over the adjoint representation of ${\rm SO}(32)$.

It is now straightforward to extract the field theory limit of the type I amplitude. Indeed, around the $SU(2)$ enhancement point, only the nearly massless vectormultiplet states survive in the field theory limit, together with the massive adjoint hypermultiplet states. Therefore, we recover the same result as in the heterotic and type II models:
\begin{equation}
	\frac{1}{(2\pi\epsilon)^2}\,\mc F(\epsilon)\bigr|_{\rm F.T.} = \sum_{k=0,\pm 1}\left[\gamma_{\hbar}(k\mu)-\gamma_{\hbar}(k\mu+ m_{\rm h}) \right] \,,
		\label{NpartFuncTypI}
\end{equation}
where $\mu=W_1-W_2$ and  $m_{\rm h}= U$ are the BPS mass parameters associated to the vector- and hyper- multiplets, respectively. As anticipated, the amplitude exactly reproduces the $\Omega$-deformed partition function of the $\mN=2^\star$ gauge theory \cite{Nekrasov:2002qd,Pestun:2007rz}.

It is possible to give a higher dimensional version of the $\Nstar$ partition function. This can be viewed as a generalisation of the Nekrasov-Okounkov formula \cite{Nekrasov:2003rj}, which describes the $\Omega$-deformed partition function of a five dimensional  $\mN=2$ gauge theory on $\mathbb R^4\times S^1$, by further including the mass deformation. To this end, we first consider the large volume limit of $T^2$, in which case the Kaluza-Klein spectrum becomes dense and the sum over momenta has to be  Poisson resummed  in \eqref{topAmplTypeI}, while keeping $U$ and $W$ fixed. By explicitly performing the Schwinger integral over $t$, one finds
\begin{equation}
	F_g \sim \frac{(2g-1)\,B_{2g}}{(2g)!} \sum_{Q\in{\rm Adj}} \left[ \mathcal E(2-g,2g-2;2U,Q\cdot W)-\frac{1}{2}\,\mathcal E(2-g,2g-2;U,Q\cdot W) \right] \,,
	\label{NO6d}
\end{equation}
where $\mathcal E(s,w;\tau,z)$ is the Kronecker-Eisenstein series of weight $w$ and is defined by
\begin{equation}
	\mathcal  E(s,w;\tau,z) = {\sum_{m,n\in\mathbb Z}}' \,(m\tau+n)^{-w}\, \frac{\tau_2^{s-\frac{w}{2}} }{|m\tau+n|^{2s-w}} \, e^{2\pi i (n\xi-m\eta)} \,,
\end{equation}
with $\xi,\eta\in \mathbb R$ and $s\in \mathbb C$. The Jacobi parameter is identified as $z=\eta+\tau \xi$. It should be stressed that \eqref{NO6d} exhibits modular covariance under the Jacobi group with weight $2g-2$. It is to be seen as a deformation of the partition function \eqref{NpartFuncTypI}, regarded as a compactification of a 6d theory on $T^2$. Considering a square $T^2$ with $U=iR_2/R_1$ and ${\rm Re}(W)=0$, and Fourier expanding the above expression, it is straightforward to obtain
\begin{equation}
	F_g =  (2\pi  R_2)^{2g-2}\,\frac{(2g-1)\,B_{2g}}{(2g)!} \sum_{Q\in{\rm Adj}}\sum_{n\in \mathbb Z}\left[ {\rm Li}_{3-2g}\left( q_U^{2n}\,e^{2\pi i Q\cdot W}\right)-{\rm Li}_{3-2g}\left(q_U^{2n+1}\,e^{2\pi i Q\cdot W}\right)\right] \,.
\end{equation}
Indeed, this is the generalisation of the Nekrasov-Okounkov formula \cite{Nekrasov:2003rj} as can be seen by re-expressing  it in the form
\begin{equation}
	F_g =   \sum_{Q\in{\rm Adj}}\sum_{n\in \mathbb Z}\left[ \gamma_g\left( \frac{Q\cdot Y +2n}{R_1} ; \beta \right)- \gamma_g\left( \frac{Q\cdot Y +2n}{R_1}+m_{\rm h} ; \beta\right) \right] \,, 
\end{equation}
where $Y$ is a real Wilson line, $\beta=2\pi R_2$ and  $m_{\rm h} = 1/R_1$.

%%%%%%%%%%%%%%%%%%%%%%%%%%%%%

\section{Conclusions}\label{Conclusion}

In this work, we have elucidated a universal structure underlying the string theory realisation of the mass deformed $\mc N=2$ gauge theory. Indeed, we have shown that freely acting orbifolds of $\mc N=4$ string compactifications, spontaneously breaking supersymmetry to $\mc N=2$, are the relevant models uplifting the $\Nstar$ gauge theory to string theory. As we have argued, the low-lying states of these models precisely match those of the mass deformed gauge theory with the adjoint mass arising from the Scherk-Schwarz deformation of the string theory.  The non-abelian group of the gauge theory, possibly including exceptional groups, can be recovered at points of enhanced symmetry in the string moduli space. These properties were explicitly demonstrated in various explicit models, with symmmetric and asymmetric orbifold actions, in type I, II and heterotic string theories, therefore supporting our proposal.

Furthermore, we have performed a non-trivial test of the correspondence by calculating topological amplitudes in the aforementioned string backgrounds. More precisely, the graviphoton amplitude $F_g$ is known to reduce, in the field theory limit, to the pure gauge theory partition function in the standard $\mc N=2$ case. Here, we have extended this property by proving, perturbatively, that the point particle limit of the graviphoton amplitude leads, in our general class of models, to the mass-deformed partition function of the supersymmetric gauge theory.

One may wonder about the fate of our correspondence at the full non-perturbative level. This is best seen from the type I or type II point of view by studying instanton corrections to the models under consideration. In addition, since in the pure $\mc N=2$ case $F_g$ is the topological string partition function, it would be interesting to understand the implications of this unveiled universality for the topological string. 
%%%%%%%%%%%%%%%%%%%%%%%%%%%%%%

\section*{Acknowledgements}

We would like to thank C. Condeescu and N. Mekareeya for useful discussions. I.F. wishes to thank the ICTP, Trieste  and
A.Z.A. would like to thank the CERN Theory Department for their warm hospitality during the accomplishment of this work.

\bibliographystyle{utphys}
\providecommand{\href}[2]{#2}\begingroup\raggedright\endgroup

\end{document}